\documentclass[twocolumn]{svjour3}  
\smartqed  

\usepackage[dvipdfmx]{color}
\usepackage{graphicx}
\usepackage[T1]{fontenc}
\usepackage{mathptmx}  
\usepackage[scaled]{helvet}  
\usepackage{courier}
\usepackage[subrefformat=parens]{subcaption}

\makeatletter
\let\cl@chapter\undefined
\makeatletter
\usepackage{amsmath,amssymb}   
\usepackage{mathtools} 
\usepackage[colorlinks=true,linkcolor=cyan,urlcolor=blue,citecolor=cyan]{hyperref}
\usepackage{cleveref}  
\crefname{equation}{Eq.}{Eqs.}%
\crefname{figure}{Fig.}{Figs.}%
\usepackage{amsbsy}
\usepackage{bm}
\usepackage{algorithmic}
\usepackage{algorithm}
\usepackage{enumerate}
\usepackage{caption,subcaption }
\usepackage[toc,page]{appendix}
\usepackage{booktabs}
\usepackage{multirow}
\usepackage{xcolor}
\usepackage{pgfplots}
\usepackage{cite}
\newcommand{\vm}{v_{\mathrm{m}}}
\newcommand{\deltas}{\delta_{\mathrm{s}}}
\newcommand{\deltap}{\delta_{\mathrm{p}}}
\newcommand{\deltasp}{\delta_{\mathrm{ps}}}
\newcommand{\degree}{^{\circ}}
\newcommand{\bt}{_\mathrm{bt}}

\newcommand{\lpp}{L_\mathrm{pp}}
\newcommand{\ld}{L_\mathrm{d}}
\newcommand{\larc}{L_\mathrm{arc}}
\newcommand{\larct}{L_\mathrm{arc}(t)}
\newcommand{\tf}{t_\mathrm{f}}
\newcommand{\lpare}[1]{\left(#1\right)} 

\crefname{table}{Table}{Tables}%
\crefname{equation}{Eq.}{Eq.}%
\crefname{figure}{Fig.}{Fig.}%
\crefname{algorithm}{Algorithm}{Algorithm}%
\crefname{section}{Section}{Section}
\creflabelformat{section}{#2#1#3}
\crefname{subsection}{Section}{Section}
\creflabelformat{subsection}{#2#1#3}
\crefname{subsubsection}{Section}{Section}
\creflabelformat{subsubsection}{#2#1#3}

\usepackage{ulem}
\providecommand{\Add}[1]{\textcolor{purple}{#1}}

\providecommand{\Erase}{\bgroup\markoverwith{\textcolor{purple}{\rule[.5ex]{2pt}{0.4pt}}}\ULon}

\usepackage{etoolbox}
\newcommand{\affaddr}[1]{#1} 
\newcommand{\affmark}[1][*]{\textsuperscript{#1}}
\usepackage[misc]{ifsym}

\journalname{Journal of Marine Science and Technology}
\begin{document}

\title{ Quantitative Evaluation of Full-Scale Ship Maneuvering Characteristics During Berthing and Unberthing

}

\author{
Agnes N. Mwange\protect\affmark[1,2*] \and 
Yoshiki Miyauchi\affmark[1] \and 
Taichi Kambara\protect\affmark[1] \and 
Hiroaki Koike\protect\affmark[1] \and
Kazuyoshi Hosogaya\protect\affmark[3] \and 
Atsuo Maki\protect\affmark[1*]
}

\authorrunning{
Agnes N. Mwange \and 
Yoshiki Miyauchi\and 
Taichi Kambara \and 
Hiroaki Koike\and
Kazuyoshi Hosogaya\and
Atsuo Maki \and 
}

\institute{
            \affaddr{\affmark[1]{Department of Naval Architecture and Ocean Engineering, Graduate School of Engineering, Osaka University, Suita, Osaka, Japan }} 
            \\\\
            \affaddr{\affmark[2]{Department of Marine Engineering and Maritime Operations, Jomo Kenyatta University of Agriculture and Technology (JKUAT), Kenya}}\\\\  
            \affaddr{\affmark[3]{Japan Hamworthy Co., Ltd.}}\\\\  
            \affmark[*]{Corresponding authors}\\
            \Letter $~$ Agnes N. Mwange\\ $~~~~~~~${mwange\_agnes\_ngina@naoe.eng.osaka-u.ac.jp \\\
            $~~~~~~~$Atsuo Maki\\ 
            $~~~~~~~$maki@naoe.eng.osaka-u.ac.jp}
}

\date{Received: date / Accepted: date}

\maketitle

\begin{abstract}
Leveraging empirical data is crucial in the development of accurate and reliable virtual models for the advancement of autonomous ship technologies and the optimization of port operations. This study presents an in-depth analysis of ship berthing and unberthing maneuvering characteristics by utilizing a comprehensive dataset encompassing the operation of a full-scale ship in diverse infrastructural and environmental conditions. Various statistical techniques and time-series analysis were employed to process and interpret the operational data. A systematic analysis was conducted on key performance variables, including approach speed, drift angles, turning motions, distance from obstacles, and actuator utilization. The results demonstrate significant discrepancies between the empirical data and the established maneuvering characteristics. These findings have the potential to significantly enhance the accuracy and reliability of conventional maneuvering models, such as the Mathematical Modeling Group (MMG) model, and improve the conditions used in captive model tests for the identification of maneuvering model parameters. Furthermore, these findings could inform the development of more robust autonomous berthing and unberthing algorithms and digital twins. 

\end{abstract}
\section{Introduction}
The research issues in ship control for the realization of automatic berthing and unberthing can be broadly divided into three categories: (i) the generation of low-speed maneuvering models, (ii) the generation of reference paths, and (iii) the formulation of control laws to follow the reference paths. 

Initially, research on maneuvering models concentrated on optimizing ship performance during long-distance voyages at cruising speeds. Among the pioneering works that laid the foundation for modern maneuvering models is the seminal study by Davidson and Schiff \cite{davidson1946turningand}. Subsequent advancements have refined ship maneuvering models from both stability and control perspectives \cite{inoue1954course, motora1955course, Nomoto1956_En} and hydrodynamic perspectives \cite{Motora1959En, Inoue1979En}. These investigations have contributed to the evolution of maneuvering models that address the hydrodynamic forces acting on the hull, rudder, and propeller, employing either integrated approaches or individual component analyses.
For instance, Abkowitz \cite{Abkowitz1964} introduced a maneuvering model that conceptualizes the ship as a rigid body, extending the representation of hydrodynamic forces through a Taylor series approximation based on state variables. In contrast, the MMG model, originally presented by Ogawa et al. \cite{Ogawa1977MMG_En} and later refined by Yasukawa et al. \cite{Yasukawa2015}, addresses the hydrodynamic forces on the hull, rudder, and propeller in a segmented fashion before integrating their interactions. It should be noted that ship dynamics and hydrodynamic forces and moments acting on a ship at low speeds differ significantly from those at cruising speeds.  For instance, the MMG model by Ogawa et al. \cite{Ogawa1977MMG_En}  defines the dimensionless yaw rate $r^{\prime}$ as $r^{\prime} = rL/U$, where $r$, $L$ and $U$ denote the yaw rate, length of the ship, and forward speed, respectively. This formulation becomes inadequate for predicting ship motions at near-zero forward speed, thereby limiting the model's applicability to low-speed scenarios.  Subsequently, to address these limitations, low-speed maneuvering models have been developed to better estimate ship dynamics during low-speed operations commonly encountered during berthing, unberthing, and navigation in confined waters. Kose et al. \cite{kose1984mathematical} introduced an alternative model in which the dimensionless $r^{\prime}$ is defined as $r^{\prime} = r/\sqrt{g/L}$ where $g$ denotes the acceleration due to gravity. Yoshimura et al.\cite{Yoshimura2009b} proposed a maneuvering model based on the concept of cross-drag flow for ocean and harbor maneuvering. 

In the context of reference path generation and control for autonomous berthing and/or unberthing, the low-speed maneuvering models previously discussed have been extensively utilized to represent ship dynamics. It is therefore essential to achieve accurate modeling of ship dynamics at low speeds in order to enhance the accuracy and reliability of path generation and control algorithms.

Developing a low-speed maneuvering model and reference path for approach and departure maneuvers\footnote{In this study, the term approach maneuvers refers to the maneuvers executed between the moment the ship begins to decelerate in preparation for port entry and the moment the ship becomes fully stationary in proximity to the berth. The berthing maneuvers encompass maneuvers executed within the period from the cessation of ship movement until the completion of the mooring process, including securing the mooring lines and the ship's final positioning alongside the berth. The departure maneuvers, which occur when leaving the port, encompass all maneuvers executed between the release of the mooring lines and the point at which the ship attains its designated cruising speed.} requires a quantitative understanding of the statistical properties of human maneuvering and the resulting ship motions.
The question of whether it is preferable to emulate human maneuvering in the automatic generation of reference paths for berthing and unberthing operations remains a topic of debate. Nevertheless, an understanding of the techniques employed by humans in such maneuvers can offer invaluable insights. The following characteristics of human maneuvering can serve as references for the automatic generation of berthing and unberthing paths:
\begin{enumerate}
    \item Speed reduction during port entry and berthing
    \item Lateral motion from the berth during unberthing
    \item Relationship between bow thruster usage and speed
    \item Effect of wind speed on the ship's path
    \item Effective distance that the ship maintains from obstacles
    \item Mean, variance, and interquartile range of the terminal positions and berthing velocity.   
\end{enumerate}

Similarly, the statistical understanding of the nature and intricacies of the actual berthing and unberthing motions is essential for the enhancement of low-speed maneuvering models as well as the development of digital twins aimed at optimizing and predicting ship performance during these operations.  Currently, there is a paucity of available quantitative information on the statistical properties of berthing and unberthing motions, which result from complex maneuvers, despite the existence of empirical knowledge. Information that could contribute to the improvement of maneuvering models and their respective captive model test conditions includes:
\begin{enumerate}
    \item From what distance and at what speed is a low-speed maneuvering model needed? What percentage of time will it occupy?
    \item Frequency of occurrence of the drift angle and propeller 4-quadrant operating conditions.
   \item Correlation between state variables such as sway velocity, $v^{\prime}$ and yaw rate, $r^{\prime}$, forward speed $u$ and the drift angle, $\beta$, or forward speed $u$ and $r$.
   \item Frequency correlation between wind and ship motion.
\end{enumerate}

Furthermore, when dynamic models are generated through system identification (SI) from navigation data of actual ships, typically involving turning and zig-zag tests that are commonly available due to the existence of sea trials data, a strong correlation is observed between the sway velocity ($v$) and yaw rate ($r$) \cite{Abkowitz1980,artyszuk2003novel}. Subsequently, when nonlinear dynamic models are estimated using these as training data, multicollinearity is induced and the coefficients are not uniquely determined  \cite{Yoon2003,luo2016parameter,wang2018quantifying}. Moreover, the range of $\beta$ and the propeller operating conditions for turning and zig-zag tests is narrower than that of berthing/unberthing motions\cite{yoshimura1990prediction,hasegawa2006study,gug2020analysis}. Therefore, if the turning and zig-zag tests are used as the training data, it would be difficult to accurately estimate the berthing/unberthing motions because the state and control inputs would be extrapolated to the training data. From this perspective, it is essential to analyze the maneuvering characteristics of the ship during berthing and unberthing.
 
\subsection{Related Research}
 The maneuvering characteristics of a ship during berthing and unberthing have been investigated from several perspectives. Honda et al.\cite{Honda1986_EN} investigated empirical speed reduction during approach maneuvers and proposed the "7-5-3" deceleration guideline. Inoue et al.\cite{Inoue2002_EN} conducted a comprehensive questionnaire survey among pilots on deceleration trends during approach maneuvers, and subsequently proposed speed reduction guidelines based on the results.  Tamaru et al.\cite{Tamaru2008_EN} measured the berthing trajectories of a coastal cement ship and developed a control strategy for approach maneuvers based on the trajectories. Sasa et al.\cite{sasa2011study} integrated questionnaire responses, field observation data, and a single logged operational dataset to simulate the influence of tidal currents on berthing operations. Hsu K.W \cite{hsu2015assessing} conducted a questionnaire survey among 16 marine pilots, each with over a decade of experience, to develop a Safety Index (SI) for assessing safety factors pertinent to in-port and berthing operations. Ozturk et al.\cite{Ozturk2019} obtained time series data of approach maneuvers using a ship-handling simulator with 20 pilots and proposed a collision risk model based on the data. 
 
 Each of these methodological approaches presents distinct advantages and limitations. Questionnaires offer the benefit of gathering extensive data across various ship types and captains at a relatively low cost. However, the data acquired may be susceptible to respondent bias and subjective interpretation. While simulators are similarly cost-effective and capable of generating data for multiple ships and ports, they are nonetheless constrained by the limitations of motion simulation and inherent modeling errors. In contrast, direct measurement of maneuvering patterns during berthing and unberthing of a specific ship provides superior data reliability. However, existing studies that have conducted statistical analyses on a sufficiently large dataset primarily focus on the final phase of berthing, evaluating parameters such as speed and impact forces
 \cite{ ISHIHATA1988, metzger2014measurement, Roubos2017}. Additionally, research on approach maneuvers has been limited to single-port operations, thereby constraining the breadth of the dataset. Moreover, previous research predominantly addresses the maneuvering techniques executed by captains, with a limited investigation of the statistical properties of berthing and unberthing maneuvering motions for the purpose of developing a comprehensive maneuvering model.

\subsection{Research Objectives and Overview}
The objective of this study is to conduct a comprehensive analysis of berthing and unberthing maneuvering data of a coastal ship that was recorded over the course of approximately one year during the operation of the ship, except for periods during which the ship was anchored. This approach facilitated the acquisition of data without incurring additional costs associated with instrumentation or specialized tests/experiments. It was posited that by aggregating data from across various ports, it would be feasible to identify generalized trends in maneuvering characteristics that are independent of port geometry. 

This study investigates empirical navigation data from a single ship to derive statistical properties of ship maneuvering during berthing and unberthing. The findings are intended to serve as a foundational reference for reference path generation, tracking control laws formulation, improvement of the model testing conditions and low-speed maneuvering models,  and the development of digital twins systems for the realization of autonomous berthing and unberthing. In addition to the preliminary study presented by the authors \cite{miyauchi20222022a}, the following improvements were made to deepen the study: $\mathrm{A}$. Addition of new data, which increased the data volume by a factor of $3$, $\mathrm{B}$. Correction of the position of the Global Navigation Satellite System (GNSS) antenna from the hull center, $\mathrm{C}$. Addition of typical measurement accuracy information of the measurement instruments, and $\mathrm{D}$. Addition of new analysis results.

\section{Notations}\label{sec:notaions}
This section provides definitions of the symbols used throughout this study. The $n$-dimensional Euclidean space is denoted by $\mathbb{R}^n$ while the set of real numbers for  $n=1$ is represented by $\mathbb{R}$.  

\section{Measurement and acquisition of ship data}
\subsection{Subject ship}
The subject ship is a coastal ship equipped with a vectwin rudder system and a controllable pitch bow thruster. The principal particulars of the ship are detailed in \cref{tab: principal particulars}. \cref{tab: data instruments} enumerates the instruments used, the physical quantities measured, and the representative measurement accuracy of each instrument. The sampling period was $ 1 \mathrm{second}$. The GNSS antenna was positioned at a distance of $ 25 \mathrm{m}$ from the midship on the bow side and $ 5 \mathrm{m}$ from the centerline on the starboard side.

\begin{figure}[htbp]
    \centering
    \includegraphics[width = \columnwidth]{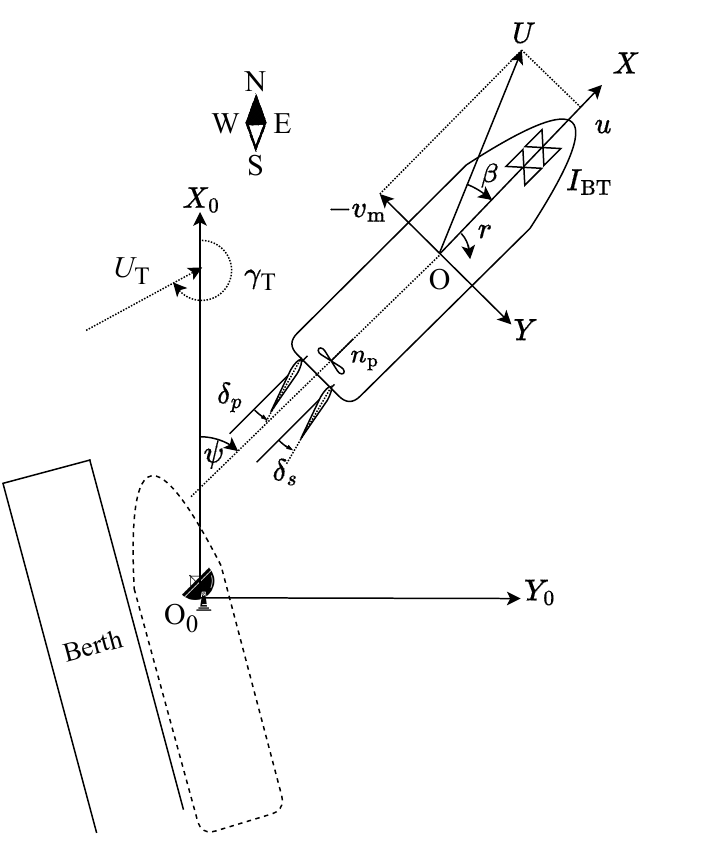}
    \caption{Coordinate systems. The origin of space-fixed $\mathrm{O}_{0}-X_{0}Y_{0}$ was set to the position of GNSS antenna at the moored position.}
    \label{fig: full scale coordinate systems}
\end{figure}

\begin{table}[htbp]
    \centering
    \caption{Principal particulars of the subject ship.  }
    \begin{tabular}{cl}
    \toprule
      Parameter   & Value \\
         \midrule
      Length ($\lpp$)   & approx. 150 m \\
      Breadth ($B$)  & approx. 25m \\
      Draft ($d$) &  approx. 8.6m \\
      \bottomrule
    \end{tabular}
    \label{tab: principal particulars}
\end{table}

\begin{table*}[htbp]
    \centering
    \caption{List of instruments and physical quantities measured. The abbreviations used in the table are defined as follows: GNSS stands for Global Navigation Satellite System, FOG stands for Fiber Optical Gyro, AIS stands for Automatic Identification System, and STBD denotes starboard.}    
    \begin{tabular}{ccc}
        \toprule
        Instruments  & Measured value & Nominal accuracy \\
        \midrule
        \multirow{6}{*}{GNSS} & Position&  10m \\
        \cmidrule(rl){2-2} %
        \cmidrule(l){3-3}
         & \multirow{2}{*}{Speed over ground (SOG)} & $<\pm0.2$ knots for SOG$<10$ knots\\
         & & $\pm2\%$  for SOG$>10$ knots\\
        \cmidrule(rl){2-2} %
        \cmidrule(l){3-3}
         & \multirow{3}{*}{Course over ground (COG)} & $\pm3.4\degree$ for $1<$SOG$<2$ knots\\ 
         &  & $\pm2\degree$ for $2<$SOG$<4$ knots  \\ 
         &  & $\pm1\degree$ for $4\text{ knots}<$SOG   \\ 
        \cmidrule(r){1-1} %
        \cmidrule(rl){2-2} %
        \cmidrule(l){3-3}       
        FOG compass & Heading and Yaw rate& static error: $\pm0.3\degree$,  $\sigma \leq 0.1\degree$\\
             \cmidrule(r){1-1} %
        \cmidrule(rl){2-2} %
        \cmidrule(l){3-3}    
       \multirow{2}{*}{Anemometer} & True wind speed&  $<\pm0.5$ m/s at wind speed <$10$ m/s\\
         & True wind direction& $<\pm5\deg$\\
     \cmidrule(r){1-1} %
        \cmidrule(rl){2-2} %
        \cmidrule(l){3-3}       
        Propeller Shaft & Shaft revolution speed& $\pm0.3\%$ \\
        Rudder system & Command and actual rudder angle& -- \\
        Bow thruster & Commanding current of thruster pitch angle& --\\
        Speed log & Speed over water& $<\pm0.1$ knots\\
        Draft meter & Draft at bow, stern and midship both PORT and STBD.&  $\pm0.5\%$\\
        AIS & Navigational status & -- \\
        \bottomrule
    \end{tabular}
    \label{tab: data instruments}
\end{table*}

\subsection{Data acquisition and dataset development}
The data was collected over two time periods: October to December 2021 and February to October 2022, which amounts to approximately one year. The data pertinent to berthing and unberthing maneuvers was extracted and analyzed in this study. The data extracted from each individual berthing or unberthing operation is referred to as ``log data''. The extracted log data was sorted into two sets, berthing and unberthing sets, and then used for the analysis. 

The criteria for extracting the log data and creating the data set are detailed below:

\begin{enumerate}[(1)]
\item For each berth, the average latitude and longitude of the GNSS antenna position was computed over an interval of approximately $ 100 \mathrm{seconds}$ during which the navigational status of the AIS indicated "Mooring". This average position was designated as the final berthing position.  Next, the average latitude and the longitude of the coordinates were transformed to the North - East space fixed coordinate system $\mathrm{O}_{0}-X_{0}Y_{0}$ with the coordinates' origin set to the final berthing position.

Since the main focus of this study is on maneuvering characteristics during berthing and unberthing, the coordinates of the midship in the $\mathrm{O}_{0}-X_{0}Y_{0}$ coordinate system are $ \mathrm{X_{ms}} = \lpare{X_\mathrm{ms}, Y_\mathrm{ms}} \in \mathbb{R}^2 $ and the limit of the Euclidean distance, $ \ld = \sqrt{X^2_\mathrm{ms} + Y^2_\mathrm{ms}}$, from the coordinates origin to the midship is $ 1.5B < \ld < 20 \lpp$. The log data was extracted after ascertaining that the data was within the set $\ld$ range, which corresponds to berthing and unberthing operations.
\item  During the measurement period, the ship visited six ports. However, data from one port, which was only visited twice, was excluded from the primary dataset due to insufficient sampling frequency. The remaining five ports were designated as Ports A, B, C, D, and E as detailed in \cref{tab: ports information}. Within Port E, operations were conducted at three separate berths, labeled Port E-1, Port E-2, and Port E-3. Due to the similarity in approach patterns at these berths, the results were combined under Port E in the analysis and subsequent sections. 
\item In the process of creating the dataset, entries with significant data loss were excluded. The final dataset consists of 82 logs of berthing operations, representing approximately 38 hours of recorded data, and 71 logs of unberthing operations, representing approximately 23 hours of recorded data. Entries with one or more missing values make up $8.0\%$ of the berthing dataset and $7.3\%$ of the unberthing dataset. 
\begin{table}[htbp]
    \centering
    \caption{Breakdown of log data for each port in the data set. }
    \begin{tabular}{p{1.1cm}|p{1.2cm}p{1.0cm}|p{1.2cm}p{1.0cm}}
    \toprule
    \multicolumn{1}{c}{Port}   & \multicolumn{2}{c}{Berthing} & \multicolumn{2}{c}{Unberthing} \\
    & No. of log data & Time (s) & No. of log data & Time (s)\\
    \midrule
        Port A & 19 & 30970       & 17 & 20668  \\
        Port B & 18 &  27581     & 18 & 23219 \\
        Port C & 9 &  16065    & 8 & 7663 \\
        Port D & 16 &  23654   & 15 &  17422 \\
        Port E-1  & 14 &  26922     & 7 &  6597 \\
        Port E-2  & 4 & 7802    &  5 &  6148 \\
        Port E-3  & 2 &  3168   &  1 &  923 \\
    \bottomrule 
         Total        & 82 & 136162  & 71 &  82640  \\
    \end{tabular}
    \label{tab: ports information}
\end{table}

\item The log data encompasses various berthing and unberthing patterns fixed for each port such as port-to-starboard orientation or port-to-port orientation as detailed in \cref{tab: berthing pattern for each port}. However, it is important to acknowledge that the dataset does not include bow and stern arrivals, a scenario typically associated with Roll-On/Roll-Off (RO-RO) ships. Additionally, there is an uneven distribution of berths across the ports analyzed, necessitating caution when generalizing the overall data distribution. Nonetheless, this does not limit the applicability of insights derived from the analysis presented in this study.  Illustrative examples of the berthing and unberthing time-series data contained in the dataset are presented in \cref{app: appendix A}.

\begin{table}[htbp]
    \centering
    \caption{Berthing patterns at each port.}
    \begin{tabular}{lc}
    \toprule
         Port   & Berthing type\\
    \midrule
        Port A & Head-in, STBD moored\\
        Port B & Head-in, PORT moored\\
        Port C & Head-out, PORT moored \\
        Port D & $90\degree$ turn to STBD, PORT moored \\
        \cmidrule(r){1-1} %
        \cmidrule(rl){2-2} %
        Port E-1 & \multirow{3}{*}{Head-out, STBD moored}\\
        Port E-2 & \\
        Port E-3 & \\
        \bottomrule 
    \end{tabular}
    \label{tab: berthing pattern for each port}
\end{table}

\item The subject ship is equipped with a vectwin rudder system and bow thruster, which enables the ship to berth and unberth without the necessity of tugboat assistance. However, the requirement for tugboat support or the use of anchors can vary based on port regulations and prevailing weather conditions.  It is difficult to obtain such event logs as digital time-series data. Therefore, the dataset utilized in this study does not differentiate between maneuvers conducted with anchor assistance or tugboat support. It is, however, known that Port B mandates the use of anchors, whereas Port E-1 requires mandatory tugboat assistance.
\end{enumerate}
\subsection{Disturbance and ship conditions}
\begin{figure}[htbp]
    \begin{minipage}[tb]{0.9\columnwidth}
        \centering
        \includegraphics[keepaspectratio, width = 0.95\hsize]{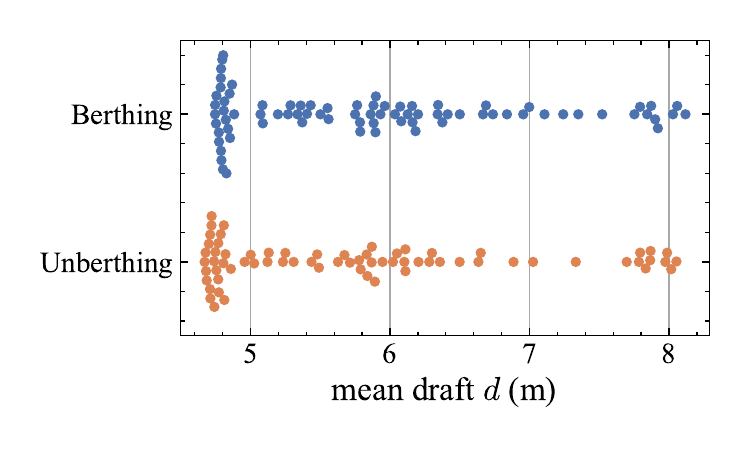}
        \subcaption{Mean draft at midship}
        \label{fig: mean draft at midship}
    \end{minipage}  
    \begin{minipage}[tb]{0.9\columnwidth}
        \centering
        \includegraphics[keepaspectratio, width = 0.95\hsize]{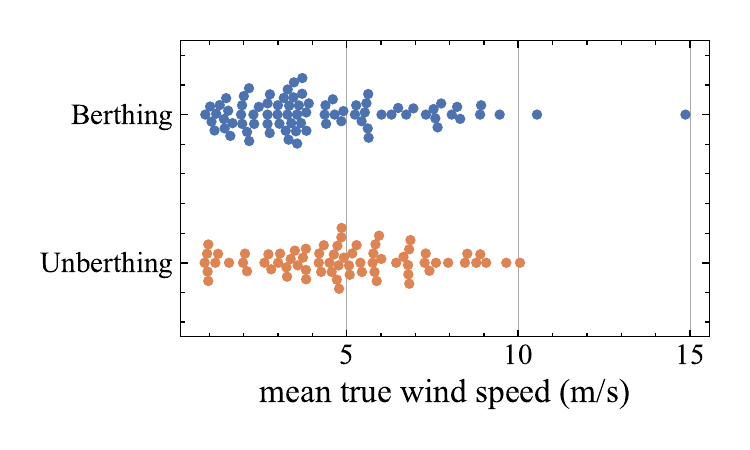}
        \subcaption{Mean true wind speed.}
        \label{fig: mean true wind speed}
    \end{minipage}
\caption{Distribution of mean draft and mean wind speed during berthing and unberthing in the data set.}
\label{fig: mean averaged draft and wind}
\end{figure}

\begin{figure*}[htbp]
  \centering
  \includegraphics[width=\textwidth]{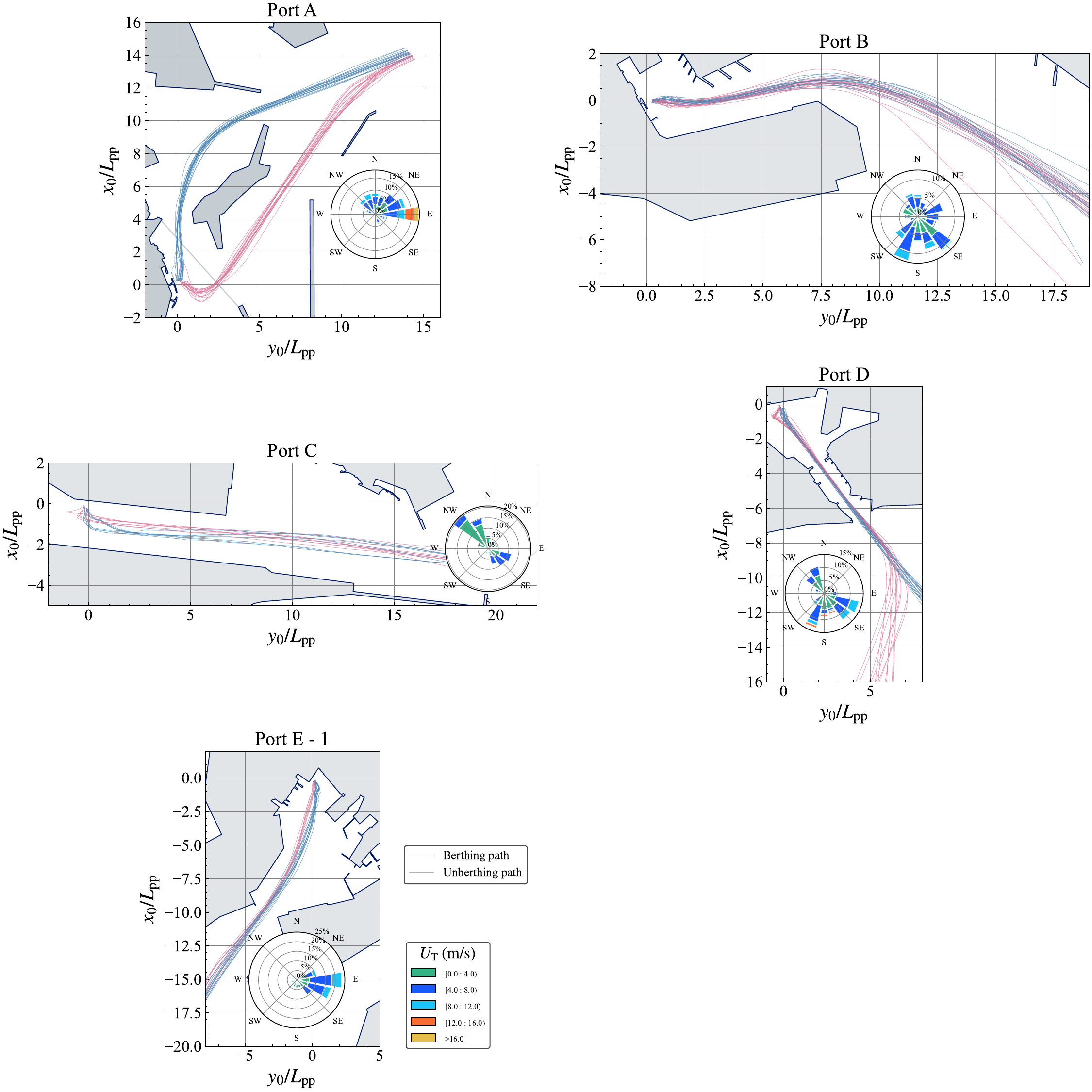}
  \caption{Windroses showing the distribution of true wind direction and speed for Ports A - E-1}
  \label{fig: windrose}
\end{figure*}

As a ship approaches the berth, it is subjected to significant environmental forces, including wind forces that exert pressure on the ship's superstructure, potentially inducing translational motions, lateral drift, and rotational motions. Simultaneously, hydrodynamic interactions with the berth and nearby structures induce complex flow patterns, including increased drag and potential suction effects, which can destabilize the ship. Within port and harbor areas, characterized by restricted maneuvering space, ensuring safe berthing and unberthing necessitates that the captain/pilot promptly implements corrective actions to mitigate these disturbances \cite{rees2021master}.  \cref{fig: windrose} illustrates the distribution of wind directions and speeds for Ports A - E-1. More details are provided in \cref{tab: wind profile }.
Moreover, since wind forces primarily act on the ship’s superstructure, the ship’s draft is a crucial parameter in moderating the effects of these forces. Accordingly, an investigation was conducted on the distribution of time-averaged wind speed, and the time-averaged ship's draft at midship for each log data as shown in \cref{fig: mean averaged draft and wind}. As illustrated in \cref{fig: mean draft at midship}, the draft is approximately $ 8.6 \mathrm{m}$ for the full load draft and there is considerable variation in the draft across the log data.
\cref{fig: mean true wind speed} indicates that the majority of the log data were collected under moderate wind speed conditions of approximately $ 10\mathrm{m/s}$, with the exception of one instance.

\begin{table}[htbp] 
\caption{Wind profile for Ports A - E-1}
\label{tab: wind profile }
     \begin{tabular}{|p{1.4cm}|p{0.9cm}|p{0.9cm}|p{0.9cm}|p{0.9cm}|p{1.0cm}|}
     \hline
     \multicolumn{1}{|c|}{}& Port A & Port B & Port C & Port D & Port E-1\\
     \hline \hline
        Mean direction, & & & & & \\
        $\Bar{\theta}_{\mathrm{w}}$ &  $40.70^\circ$ &$160.9^\circ$ & $4.700^\circ$ & $157.5^\circ$& $105.1^\circ$\\
        \hline
        Mean resultant length, & & & & & \\
        $\bar{R}_{\mathrm{w}}$ & 0.458 &0.192 & 0.195& 0.268 & 0.505  \\
        \hline
        Variance,& & & & & \\ $s_{\mathrm{w}}$ & 0.542 &0.808 & 0.805& 0.732 & 0.495\\
        \hline
        Angular standard & & & & & \\
        deviation & 1.040 &1.271 &1.269 &1.210 &0.995\\
        \hline
        Circular standard & & & & & \\ deviation  & 0.823  & 1.197 & 1.191 & 1.070 & 0.771\\
    \hline 
    \end{tabular}
 \end{table}
 
\section{Analysis Results}

\subsection{Statistical properties of berthing and unberthing motions }
This section presents an analysis of the statistical properties of ship motions during berthing and unberthing operations based on the created dataset. In this study, ship speed ($ U$), drift angle ($\beta$), and yaw rate ($ r$) were identified as the primary variables in analyzing maneuvering motions. Subsequently, their variation and correlation with other variables were systematically analyzed. It is important to note that the scatter plots (\cref{fig:beta_rstar,fig:beta_r_prime,fig:dist_bt,fig:u_bt}) shown in this section were generated using a dataset resampled at $ 0.1 \mathrm{Hz}$ to enhance image clarity. 

\subsubsection{Cumulative path length}
In the context of berthing and unberthing maneuvers, the cumulative path length ($\larc$) refers to the total distance traversed by the ship along its trajectory to or from the berthing point. This length can also be expressed in terms of the Euclidean distance between the ship's position and the target berthing point ($\ld$) and can be used to analyze properties such as ship speed at the beginning of approach maneuvers. However, if the ship is berthed with an outbound orientation, and makes a 180-degree turn, it is easy to imagine that multiple points along its path could take the same $\ld$.  Therefore, this section presents the comparison between $\ld $ and the path length $\larc$ for each log data. To facilitate this comparison, for each log data, the cumulative distance $\larct$ at a given time $t$ refers to the arc length of the path traversed by the ship from a time $ t = t_0$ to time $\tf$ for berthing, and from time $t = t_0 $ to the time $t$ at the end of the log data entry for unberthing. $\larct$ is expressed as shown in \cref{eqn: L_arc} :

\begin{multline}
  \larct =\\
     \begin{cases}
      {\displaystyle \sum_{i=t_{0}}^{t} \sqrt{\big(x(i)-x(i-1)\big)^2+\big(y(i)-y(i-1)\big)^2} + \ld(t_{0})}\; \\ \text{for: Unberthing}\\\\
      {\displaystyle \sum_{i=t}^{t_{\mathrm{f}}} \sqrt{\big(x(i)-x(i-1)\big)^2+\big(y(i)-y(i-1)\big)^2} + \ld(t_{\mathrm{f}})}\; \\ \text{for: Berthing}\\
   \end{cases}
   \label{eqn: L_arc}
 \end{multline}
 
\begin{figure}[htbp]
    \centering
    \includegraphics[width = \columnwidth]{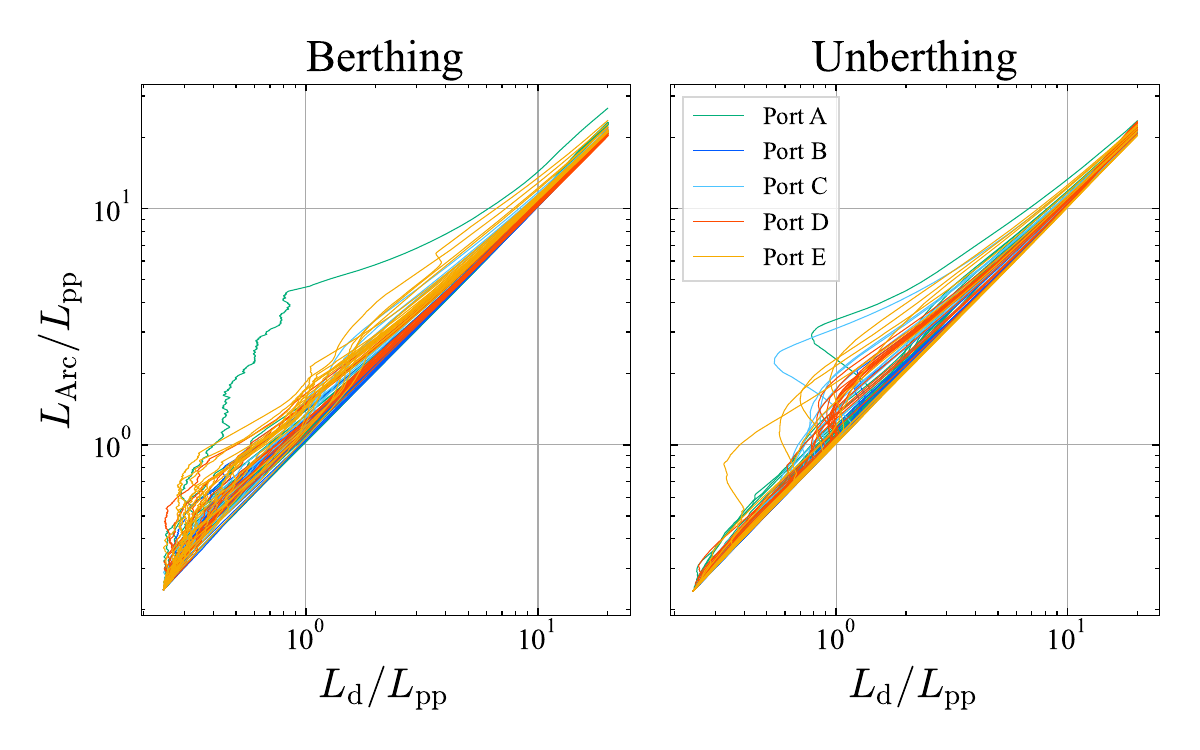}
    \caption{Relationship between the Euclidean distance $\ld$ and arc length of trajectory $\larc$. Each line represents the individual log data.}
    \label{fig: relationship between L_D and L_arc}
\end{figure}

\cref{fig: relationship between L_D and L_arc} shows the relationship between $\ld$ and $\larc$ for each log data. For both berthing and unberthing, there is a region within the range $0 < \ld / \lpp < 2$ where a single value of $\ld$ corresponds to multiple values of $\larc$. This indicates that within the range $ 0< \ld/\lpp < 2$, there exist some points that are not uniquely defined by $\ld$ due to the curvature of the berthing or unberthing path. Nevertheless, in the region farther from the berth, there is no significant difference between $\ld$ and $\larc$, indicating that both metrics can be used interchangeably to analyze the ship’s maneuvering characteristics. Henceforth, $\ld$ will be used to express the distance of the ship from the berth.

\subsubsection{Speed Reduction}
Speed reduction is an important parameter in planning the berthing process. The captain/pilot is required to maneuver the ship to stop parallel to the berth at the target berthing point within a sufficiently short time, with appropriate deceleration to avoid overrun, while maintaining a speed that allows the ship to resist external disturbances\cite{international2003colreg,guard2010new}. Consequently, research on deceleration trends has been conducted, and guidelines have been proposed \cite{Inoue2002_EN, KiyoshiHara1976,inoue1991assessment}. In the context of autonomous navigation, path generation algorithms play a crucial role in the realization of autonomous berthing and unberthing. While it has been demonstrated that appropriate speed reduction can significantly reduce the likelihood of accidents  \cite{chang2019impact}, the majority of path generation and control algorithms primarily focus on the initial and final berthing speeds   \cite{djouani1995minimum, LIAO201947, martinsen2020optimization, maki2021application, rachman2022warm, YUAN2023113964}. Notably, only $8\%$ of existing autonomous navigation incorporates ship speed as part of the objective function  \cite{OZTURK2022111010}. Ship speed during berthing and unberthing has been considered from various perspectives. For instance, Miyauchi et al. utilized ship speed during approach and departure maneuvers to define spatial constraints for collision avoidance \cite{miyauchi2022optimization}. Additionally, the significance of speed reduction has been demonstrated in the context of trajectory optimization\cite{mwange2023online} and its role in automatic control \cite{hasegawa1993mathematical, rachman2023experimental}. 

\begin{figure}[htbp]
    \centering
    \begin{minipage}[b]{\columnwidth}
        \centering
        \includegraphics[width=0.95\columnwidth]{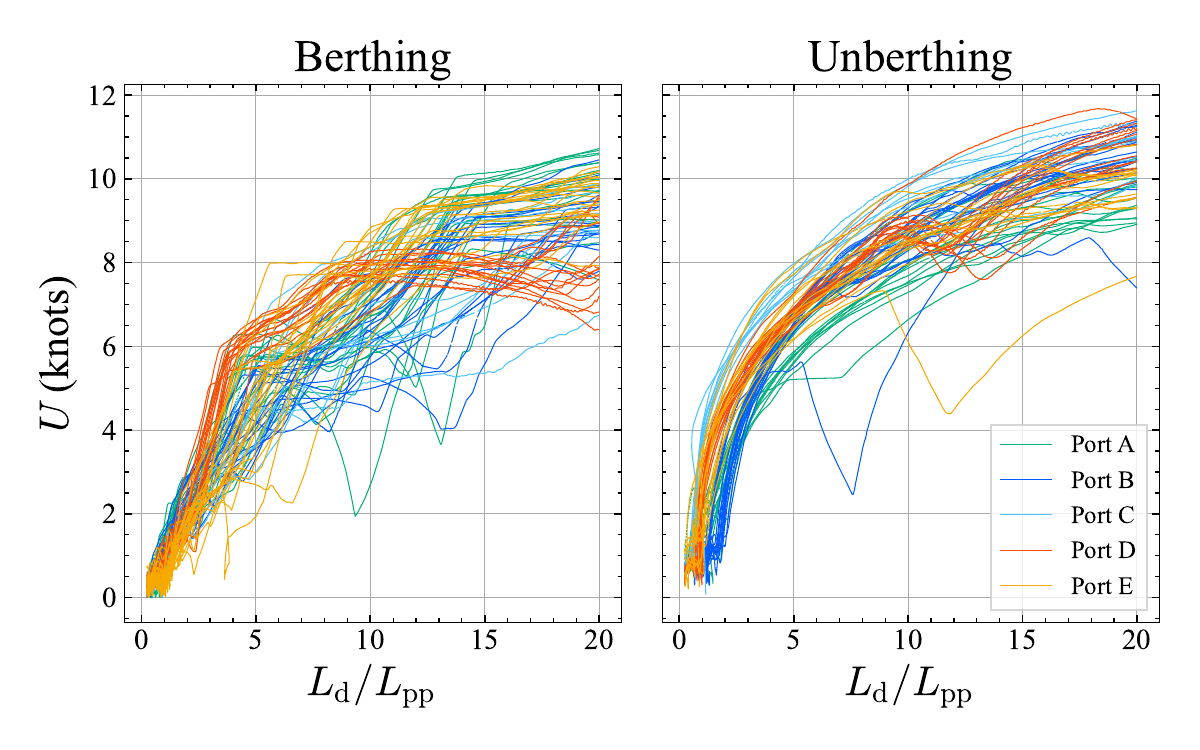}
        \subcaption{Linear scale.}
    \end{minipage}\\
    \begin{minipage}[b]{\columnwidth}
        \centering
        \includegraphics[width = 0.95\columnwidth]{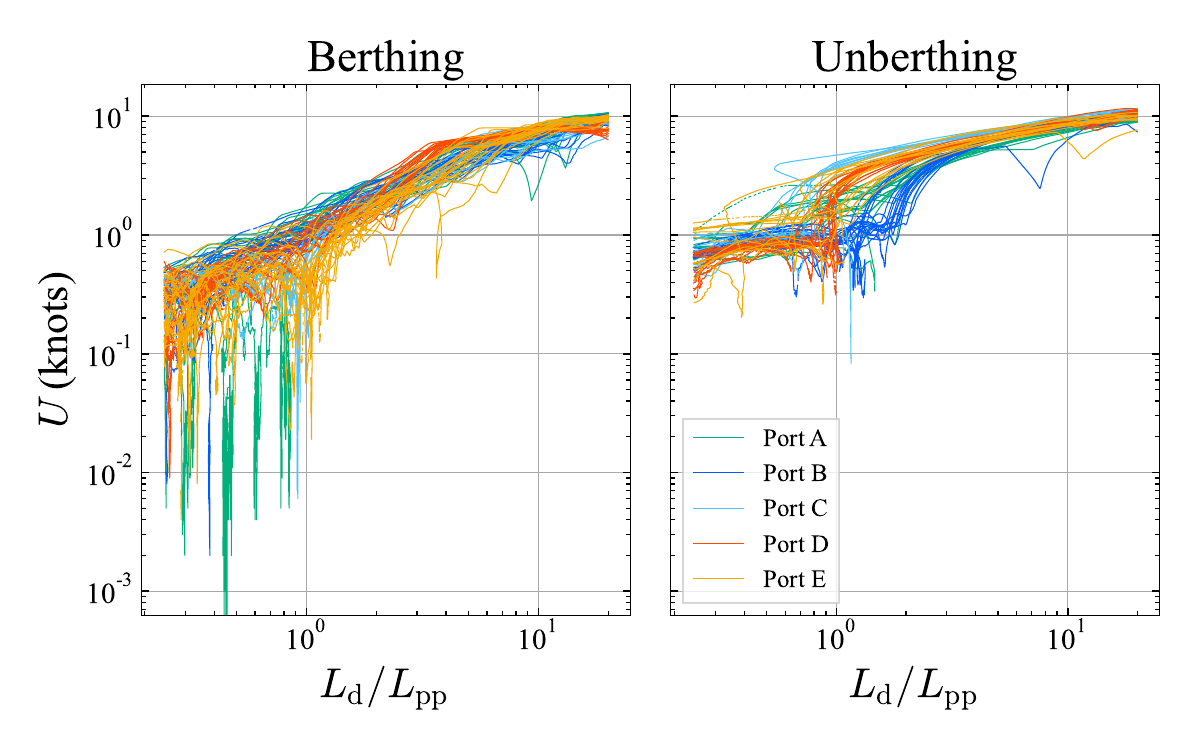}
        \subcaption{Log scale.}
    \end{minipage}
    \caption{Speed over ground $U$ during berthing and unberthing.}
    \label{fig: dist_hist_U_log}
\end{figure}

In this study, the deceleration patterns were analyzed by evaluating the relationship between the speed over ground, $U = \sqrt{ u^2 + \vm ^ 2}$ and the distance $\ld$ for each berthing and unberthing log data as illustrated in \cref{fig: dist_hist_U_log}. 

\begin{figure}[htbp]
    \centering
    \includegraphics[keepaspectratio, width=0.95\columnwidth]{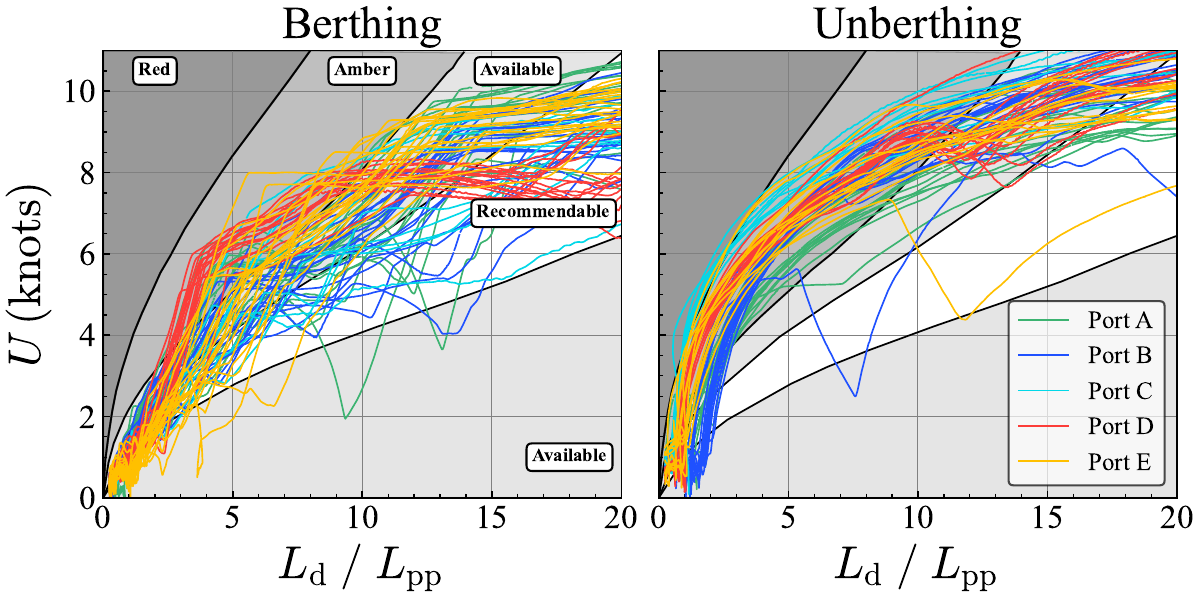}
   \caption{Comparison of the berthing/unberthing speed with the speed reduction guidelines proposed by Inoue et al.\cite{Inoue2002_EN}}
    \label{fig: comparison with Inoue guidelines}
\end{figure}

Additionally, the deceleration patterns were evaluated in comparison with speed reduction guidelines proposed by Inoue et al.\cite{Inoue2002_EN} as shown in \cref{fig: comparison with Inoue guidelines}. In essence, the guidelines establish a correlation between the level of safety and two key factors: the braking force applied and the distance from the berth. In the "Red" region, even with the application of the maximum braking power, the ship cannot reach zero speed before reaching the berth. In contrast, in the "Amber" region, the ship is capable of stopping, although there is an inherent risk of losing control. The "Available" regions are considered to be safe for operation and do not necessitate the application of maximum braking power. The "Recommendable" region represents the operational area within which the majority of captains typically operate.  As illustrated in \cref{fig: comparison with Inoue guidelines}, despite the ship's excellent braking power and maneuverability, it was operated mostly in the upper `Available' and `Recommendable' regions during berthing. Moreover, in the vicinity of the berth, the ship was primarily operated within the `Recommendable' or lower `Available' regions. On the other hand,  during unberthing, the ship was predominantly operated in the `Amber' and upper `Available' regions. These findings reinforce the applicability of these guidelines \cite{Inoue2002_EN}, particularly when planning the berthing process. Additionally, these findings highlight the need for the formulation of guidelines tailored to the unberthing.

\subsubsection{Distance from obstacles}
\begin{figure}[htbp]
 \begin{subfigure}[b]{0.5\textwidth}
    \centering    \includegraphics[keepaspectratio, width=0.95\hsize]{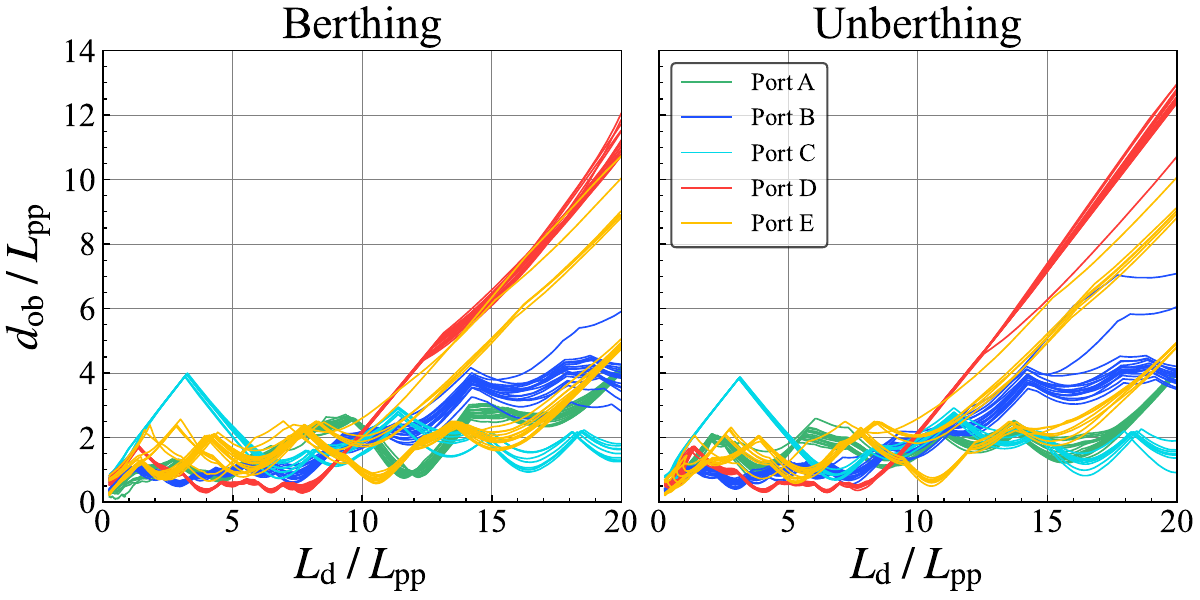} 
        \caption{}
        \label{fig: dist from port wall vs ld}
\end{subfigure} 
\begin{subfigure}[b]{0.5\textwidth}
    \centering
        \includegraphics[keepaspectratio, width=0.95\hsize]{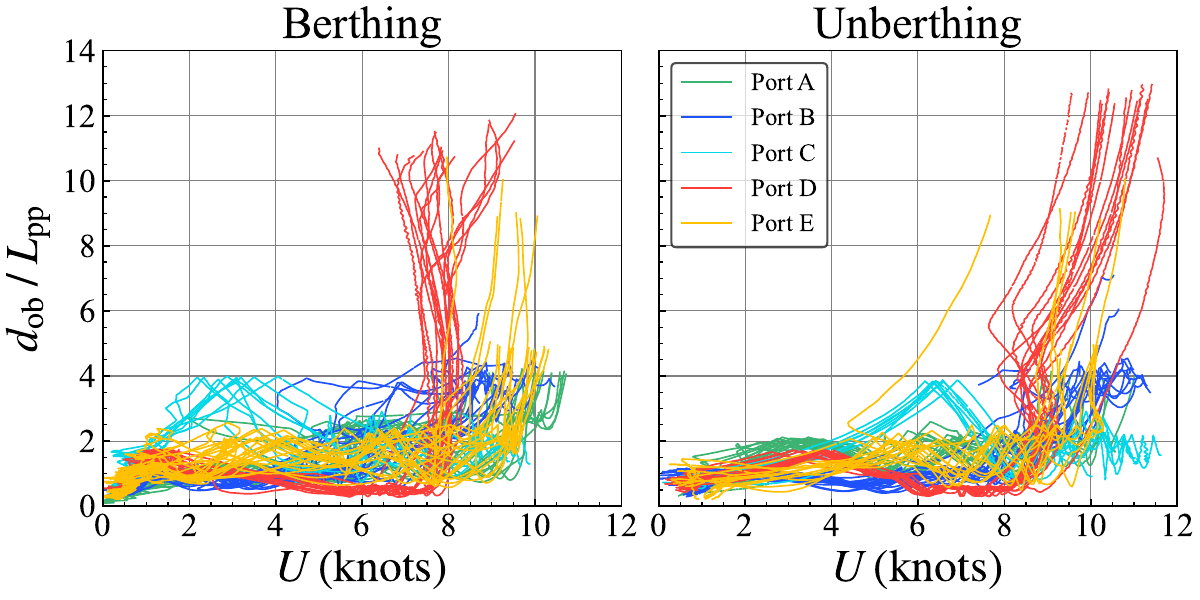} 
        \caption{}
        \label{fig: dist from port wall vs U}
\end{subfigure} 
\caption{Relationship between: \subref{fig: dist from port wall vs ld} the distance from the nearest port wall, $d_{\mathrm{ob}}$, and the Euclidean distance, $L_{\mathrm{d}}$, and \subref{fig: dist from port wall vs U} the distance from the nearest port wall, $d_{\mathrm{ob}}$, and the speed over ground, $U$.}
\label{fig: distance from obstacles}
\end{figure}

Maintaining a safe distance from both static and dynamic obstacles is fundamental to ensuring collision avoidance and operation safety by allowing captains/pilots sufficient space and time to respond to internal or external disturbances. This distance is also referred to as the distance of the closest point of approach (DCPA) and in previous research, the distance has been estimated analytically and/or empirically as a distance or region (ship domain) \cite{hansen2013empirical,wang2016empirically,silveira2022method,zhou2022online}. Moreover, in the context of autonomous berthing and unberthing the distance has been used to define spatial constraints in trajectory optimization algorithms \cite{rachman2022warm,miyauchi2022optimization,han2022potential}. It is evident that empirical knowledge of these distances is indispensable for the development of reliable autonomous berthing and unberthing algorithms, as it guarantees precise collision prediction and avoidance. Furthermore, this information is also crucial for advancing digital twins, which ultimately contributes to safer and more efficient autonomous berthing and unberthing. \cref{fig: distance from obstacles} illustrates the minimum distance from static obstacles (port walls) that was maintained by the ship with respect to distance from the berth, $\ld$, and speed over ground, $U$.

\subsubsection{Drift angle $\beta$ }\label{sec:beta}     
The standard MMG model \cite{Ogawa1978b} widely used for estimating motion assumes a small drift angle $\beta$. However, when the drift angle becomes large, it is desirable to use an MMG model that assumes low-speed maneuvering (low-speed maneuvering model) \cite{Yasukawa2015}. Accordingly, a thorough comprehension of the characteristics of $\beta$ during the berthing and unberthing operations of a full-scale ship is crucial in the application or refinement of MMG models. This section addresses the following: 
\begin{enumerate}
    \item Relationship between $\beta$ and $\ld$
    \item Relationship between $\beta$ and $U$
    \item Relationship between $\beta$ and the rudder angles, $\deltasp$
\end{enumerate}

\begin{figure}[htbp]
    \begin{subfigure}[b]{0.5\textwidth}
    \centering
    \centering
    \includegraphics[keepaspectratio, width=0.95\columnwidth]{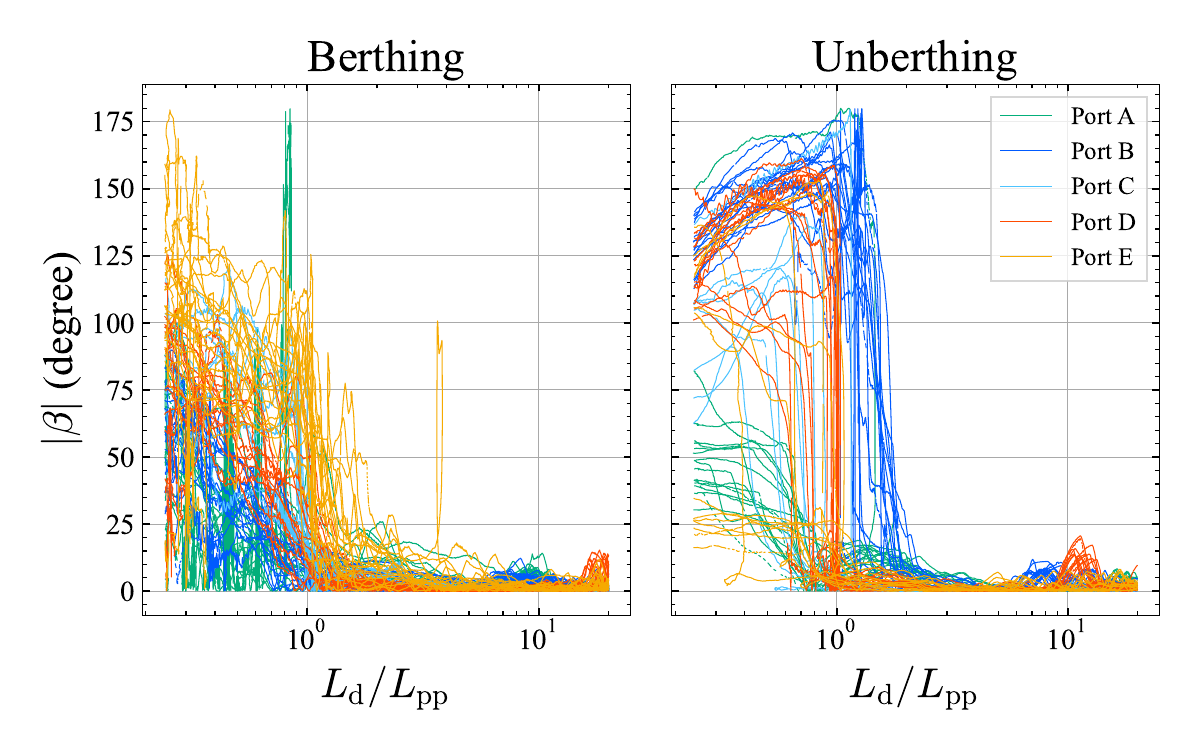}
    \end{subfigure}
    
    \begin{subfigure}[b]{0.5\textwidth}
    \centering
    \includegraphics[keepaspectratio, width=0.95\columnwidth]{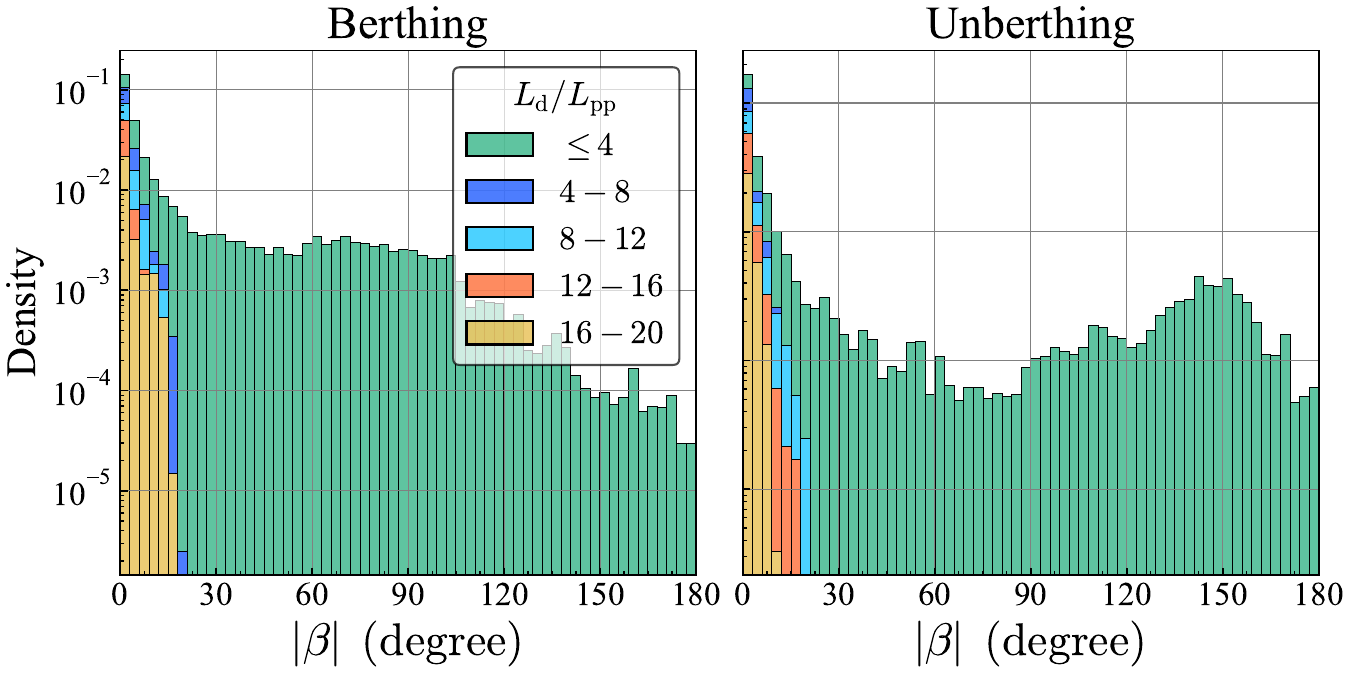}
    \caption{}
    \label{fig: beta_ld_hist_all}
    \end{subfigure}
    
    
    \caption{Distribution of the drift angle, $|\beta|$ with respect to $\ld$ during berthing and unberthing.}
    \label{fig:dist_hist_beta}
\end{figure}

\begin{figure}[htbp]
    \centering
    \includegraphics[keepaspectratio, width=0.9\hsize, trim=100 160 630 0, clip]{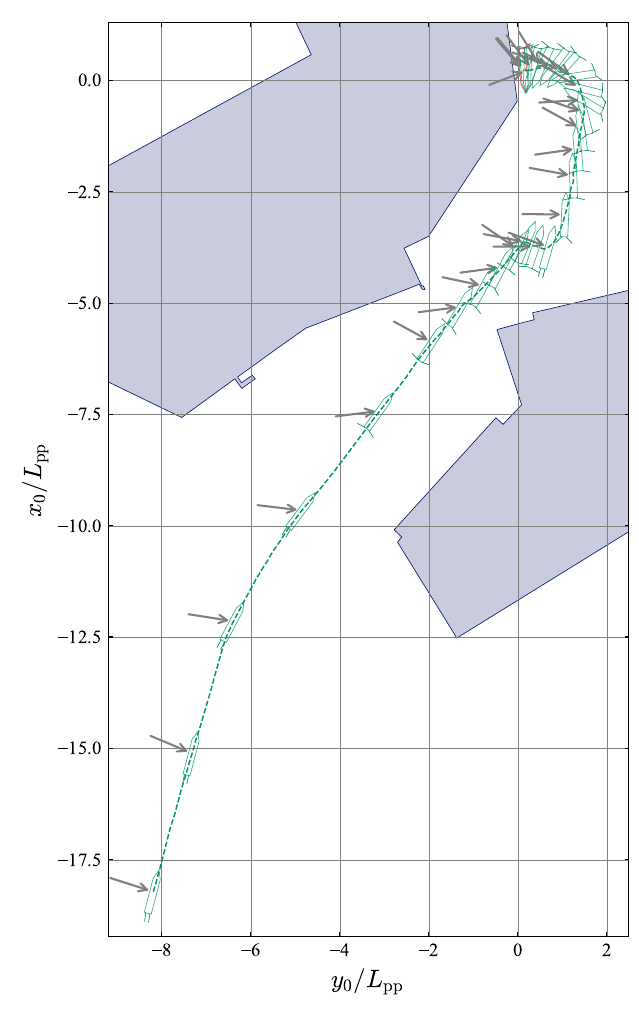}
    \caption{Trajectory of a log data with anomaly drift angle at $\ld/\lpp\approx4$.}
    \label{fig:anomaly_beta}
\end{figure}

\begin{figure}[htbp]
    \centering
    \includegraphics[keepaspectratio, width=0.95\hsize]{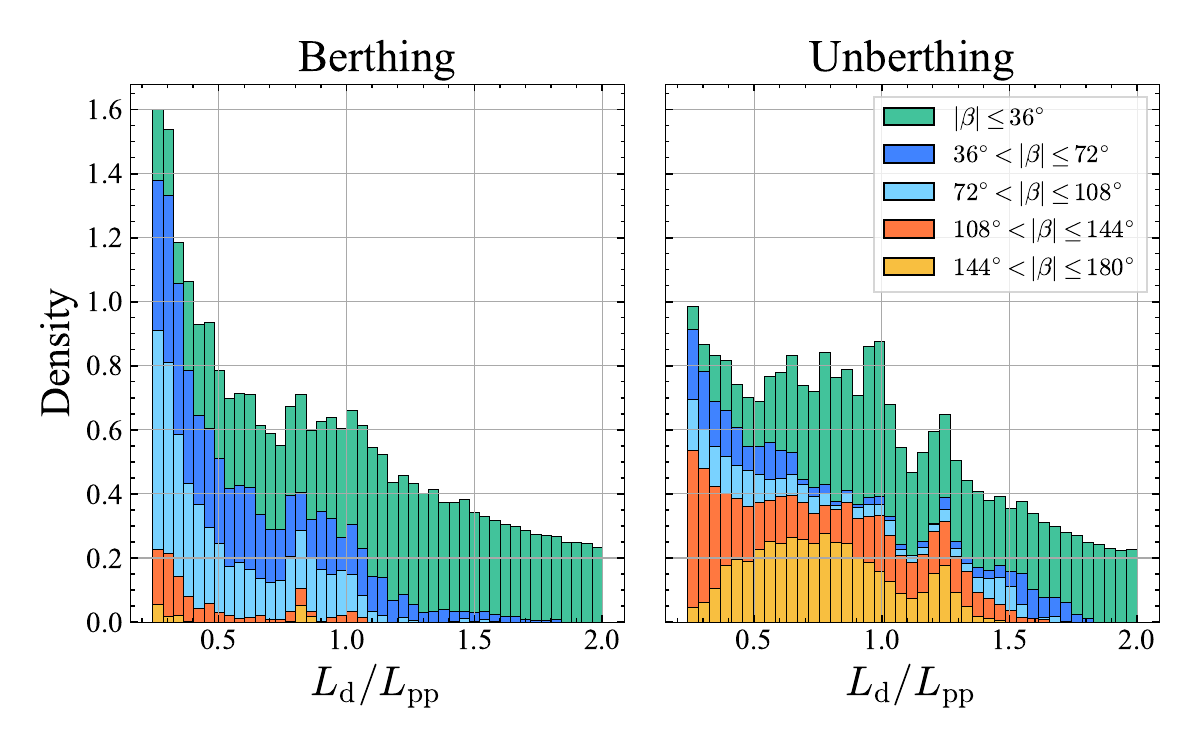}
    \caption{Distribution of $\beta$ across the $\ld / \lpp \leq 2$ range. Histograms are normalized so that the sum of their areas equals 1 in each figure.}
    \label{fig:beta_cat_2L}
\end{figure}

First, the relationship between $|\beta|$ and the distance $\ld$ was analyzed as shown in \cref{fig:dist_hist_beta}. With the exception of a berthing log data at Port E, deviations from the range of drift angle $|\beta|<30\degree$ assumed in the standard MMG model were predominantly observed in regions proximal to the berth, $\ld\leq2\lpp$. \cref{fig:anomaly_beta} shows the exceptional case of the log data with $|\beta| < 30^\circ$ for $\ld < 2\lpp$. The occurrence of this outlier is assumed to result from the ship being momentarily stopped at a distance approximately $\ld/\lpp \approx 4 $ from the berthing point before the rudder was switched to parallel steering mode with forward propulsion to counteract wind disturbances (the steering modes are detailed in \cref{sec:rudderangle}). As illustrated in \cref{fig:anomaly_beta} the ship appears to have been strongly drifting to the starboard side because of shore westerly winds. Henceforth, we designated $\ld \leq 2.0\lpp$ as the range associated with large drift angles and analyzed the distribution of $|\beta|$ within this range. In the dataset, approximately $43.5\%$ berthing data and $40\%$ unberthing data fall within the $\ld \leq 2.0\lpp$ range.
\cref{fig:beta_cat_2L} illustrates the distribution of $|\beta|$ in the range of the large drift angles, divided into five categories. Additionally, as illustrated in \cref{fig:beta_cat_2L}, it is evident that even in the berthing scenarios that require a $180\degree$ turn to berth the ship with an outbound orientation,  $72\degree<|\beta|$ is observed within  $\ld/\lpp<1$, which corresponds to a backward or lateral movement during berthing. On the other hand, unberthing operations included backward and lateral movements up to a longer $\ld / \lpp <1.5$, indicating that large drift angles must be considered during both berthing and unberthing.

\begin{figure}[htbp]
    \begin{subfigure}[b]{0.5\textwidth}
    \centering
        \includegraphics[keepaspectratio, width=0.95\hsize]{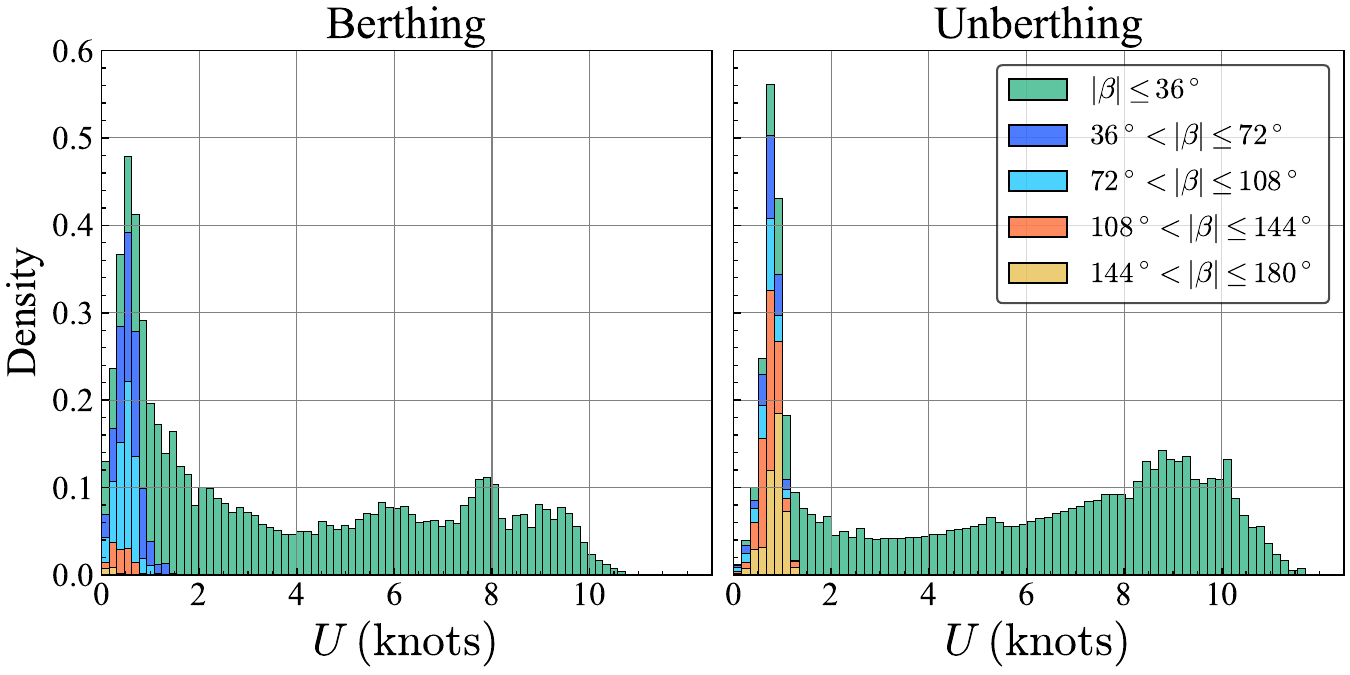}
        \caption{}
        \label{fig: beta_U_hist_all}
    \end{subfigure}
    \begin{subfigure}[b]{0.5\textwidth}
    \centering
        \includegraphics[keepaspectratio, width=0.95\hsize]{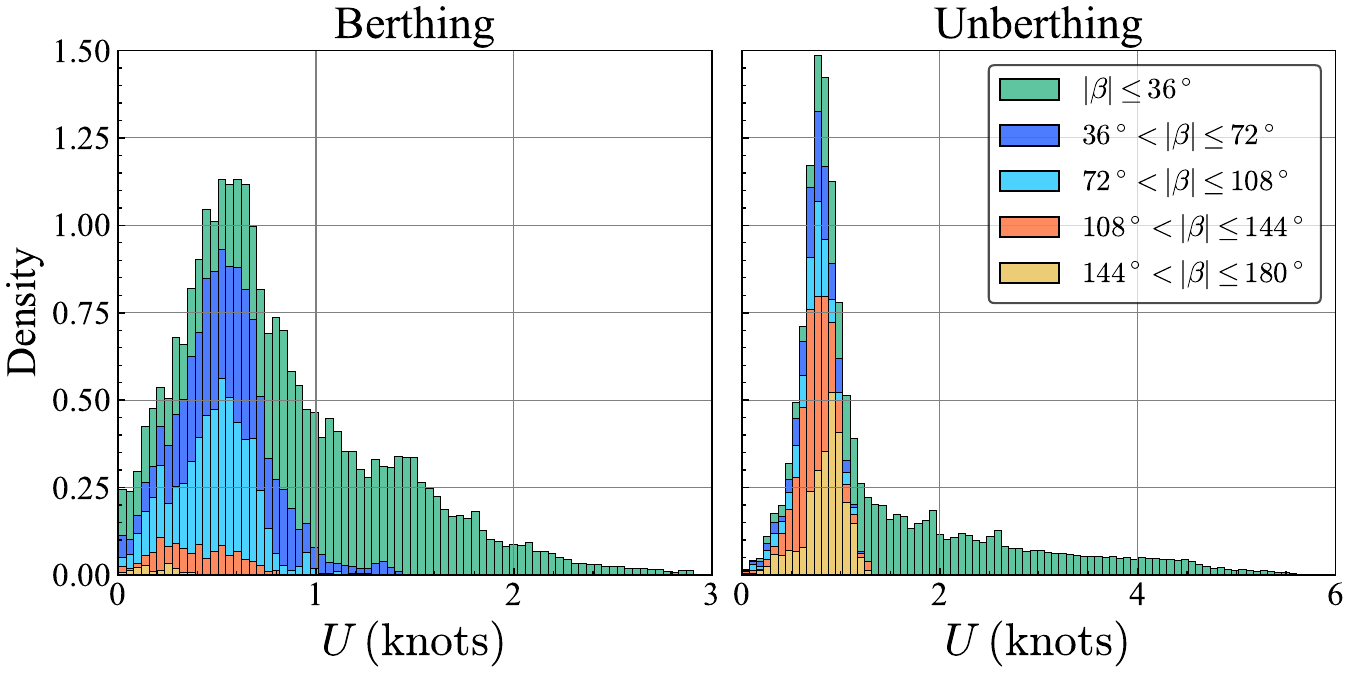} 
        \caption{}
        \label{fig: beta_U_hist_ld<2}
    \end{subfigure} 
    \caption{Distribution of $\beta$ across \subref{fig: beta_U_hist_all} the entire $U$ range and \subref{fig: beta_U_hist_ld<2} $U$ in $\ld\leq 2$ range. Histograms are normalized such that, the sum of their areas equals 1 in each figure.}
    \label{fig:beta_cat_u}
\end{figure}

Next, the relationship between $\beta$ and $U$ in the range $\ld / \lpp \leq 2$ was analyzed. As illustrated in \cref{fig: beta_U_hist_ld<2}, large drift angles, $|\beta| > 36\degree$, occur at speeds lower than $ 1$ knot when berthing and $1.5$ knots when unberthing, and when moving laterally or sternways at low speed. Therefore, in the context of low-speed maneuvering models,  it is sufficient to consider very low speeds such as $2$ knots or less, as the range below which large drift angles occur.

\begin{figure}[htbp]
    \begin{subfigure}[b]{0.5\textwidth}
    \centering
        \includegraphics[keepaspectratio, width=0.85\hsize]{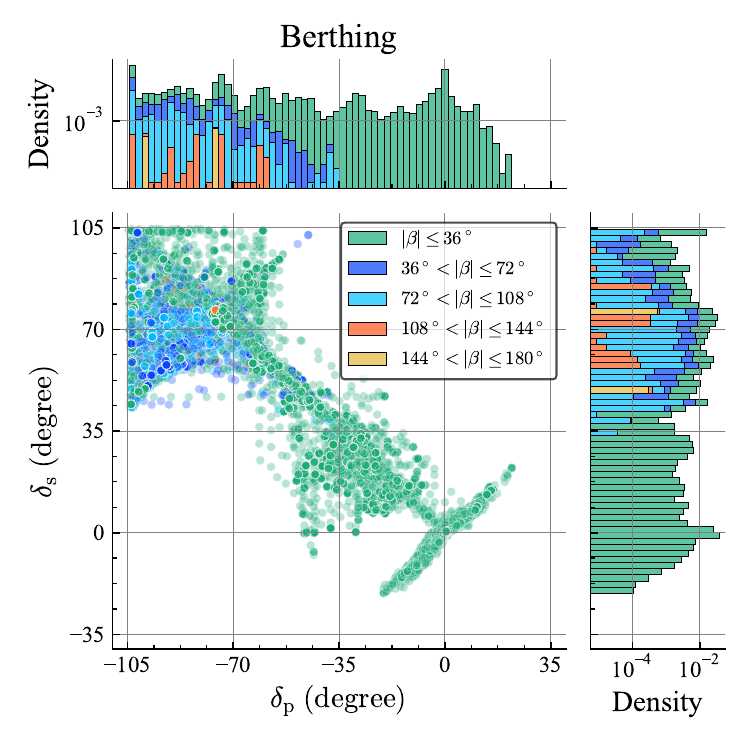}
        \caption{}
        \label{fig: berthing rudder vs beta}
    \end{subfigure}
    \begin{subfigure}[b]{0.5\textwidth}
    \centering
        \includegraphics[keepaspectratio, width=0.85\hsize]{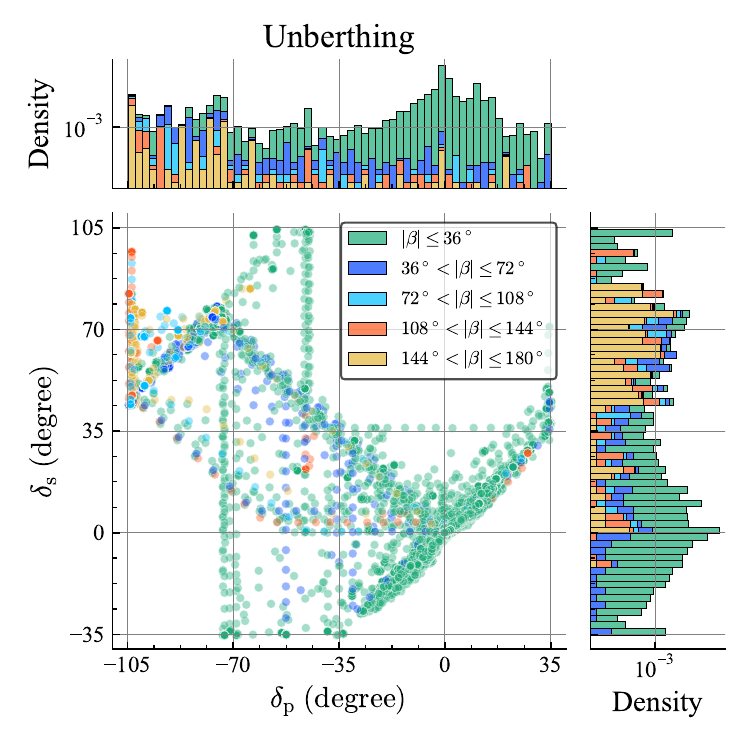} 
        \caption{}
        \label{fig: unberthing rudder vs beta}
    \end{subfigure} 
    \caption{Relationship between the rudder angles and the drift angle during \subref{fig: berthing rudder vs beta} berthing and \subref{fig: unberthing rudder vs beta} unberthing. $\deltap$ and $\deltas$ refer to port and starboard rudder angles, respectively. The histograms are normalized such that the sum of the bar areas equals 1 and vertical axes are in log scale. }
    \label{fig: beta_rudder}
\end{figure}

Finally, the relationship between $|\beta|$ and the rudder angles, $\delta_\mathrm{s, p}$, was analyzed. For a standard single screw, single rudder ship, the relationship between drift angle and rudder angle is such that the application of a rudder angle induces a corresponding drift angle. This drift angle influences the cross-flow dynamics at the stern, thereby resulting in a reduction of effective rudder incidence angle \cite{molland2011maritime}. As illustrated in \cref{fig: berthing rudder vs beta} and \cref{fig: unberthing rudder vs beta}, large drift angles, $|\beta| > 36\degree$, are more prevalent during unberthing than berthing. This can be attributed to the steering modes used, (discussed in detail in \cref{sec:rudderangle}) which are characterized by different magnitudes of the rudder angles. Difference between the absolute magnitude of the starboard and port rudder angles results in a drift angle, uneven flows on the rudder, and different rudders' normal forces and moments \cite{kang2008mathematical,hasegawa2006study}. This highlights the need for in-depth research to develop a mathematical correlation between the vectwin rudder system steering modes and drift angles.

\subsubsection{Angular turning velocity, $r$}\label{sec:r_yaw}
In this section, the relationship between $\beta$ and $r$ was analyzed. In the context of dynamic models, the non-dimensionalization $r^{\prime} = r\lpp/U$ is often used. However, since $U$ fluctuates during berthing and unberthing, the corresponding dimensionless $r^{\prime} = r\lpp/U$ becomes significantly large. On the other hand, in constrained model tests for system identification of model coefficients, the measurement range of $r^{\prime}$ is considerably narrower compared to that of the standard MMG method, which is approximately $|r^{\prime}| \leq 0.8$ \cite{Yasukawa2015}. For the models assuming low-speed maneuvering, the range is about $|r^{\prime}| \leq  2$ \cite{Yoshimura2009b, Yumuro1988_EN}.  Therefore, this study investigated the potential range of $r^{\prime}$ values encountered during the berthing and unberthing of a full-scale ship. To achieve this, the yaw rate $ r_{t}$ at a given time $t$ of berthing and unberthing was non-dimensionalized using $U = U_{t}$ and $\lpp$ at the same time.

\begin{equation}
    r^{\prime}_{t}=r_{t}\lpp/U_{t}
\end{equation}

\begin{figure}[htbp]
    \begin{subfigure}[b]{0.5\textwidth}
        \centering
        \includegraphics[keepaspectratio, width = 0.95\hsize]{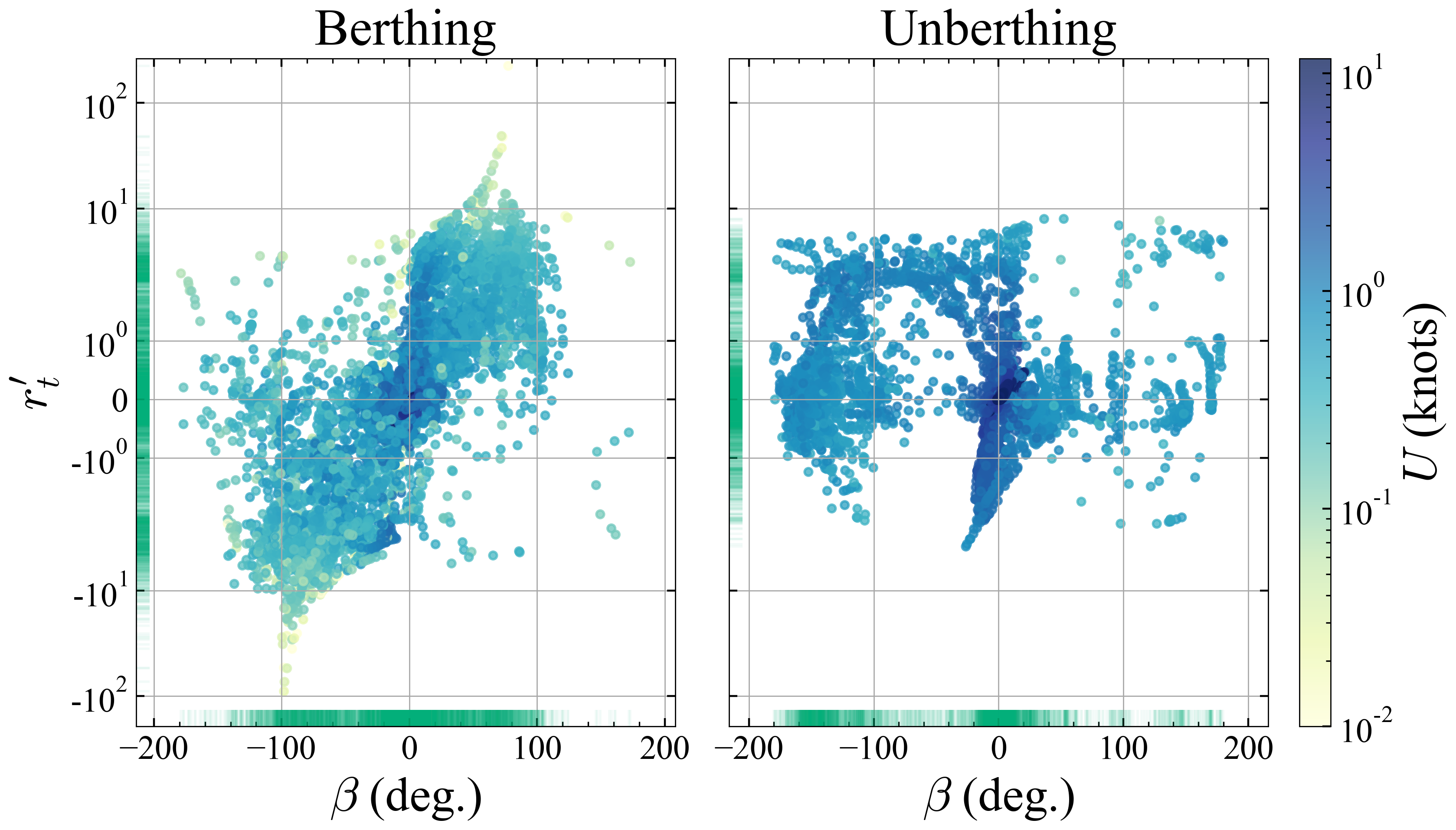}
    \caption{}
    \label{fig: beta_r_prime_scatter}
    \end{subfigure}
    \begin{subfigure}[b]{0.5\textwidth}
        \centering
        \includegraphics[keepaspectratio, width = 0.95\hsize]{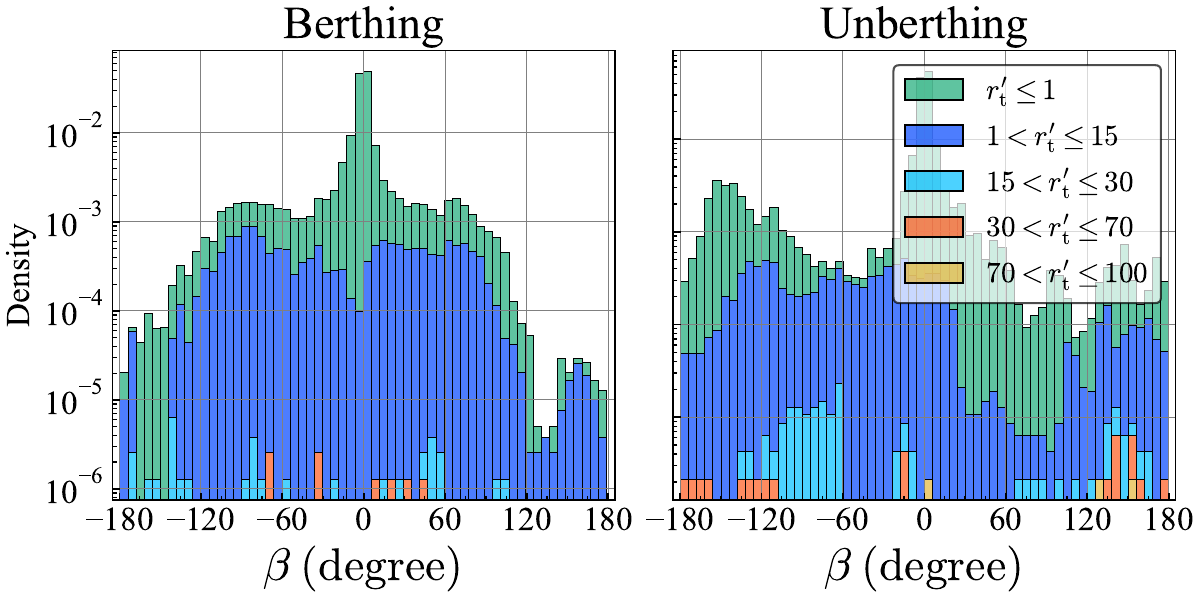}
    \caption{}
    \label{fig:beta_r_prime_histogram}
    \end{subfigure}
\caption{Correlation between $\beta$ and $r^{\prime}_{t}$. The histogram is normalized such that, the total area of the bars equals 1 for each figure.}
\label{fig:beta_r_prime}
\end{figure}

The relationship between $r^{\prime}, \; |\beta|$ and $\;U$ is shown in \cref{fig:beta_r_prime}. It is evident that, for both berthing and unberthing, points where $| r^{\prime}| > 1$ are distributed over the entire $|\beta|$ range. Additionally, during berthing, there are instances where, for large forward drift angles, $50\degree\leq|\beta|\leq80\degree$, the yaw rate falls within the range $1\times10^{1}\leq|r^{\prime }_{t}|\leq1\times10^{2}$. This range significantly exceeds the limits typically observed in conventional captive model tests, where $|r^{\prime}|\leq2$. Furthermore, the speed range corresponding to $|r^{\prime}_{t}|>2$ is approximately $0.1$ to $1$ knots for berthing and approximately $0.2$ to $2$ knots for unberthing. The range of $r^{\prime}_{t}$ during berthing and unberthing observed in this study extends beyond the measurement range of conventional captive model tests. Consequently, it is anticipated that the performance of the model within the deviated range of $r^{\prime}_{t}$ will be verified, thereby necessitating potential revisions to the model test conditions and subsequent enhancements to the maneuvering models.

\begin{figure}[htbp]
    \begin{minipage}[tb]{\linewidth}
    \centering    \includegraphics[keepaspectratio, width = 0.95\hsize]{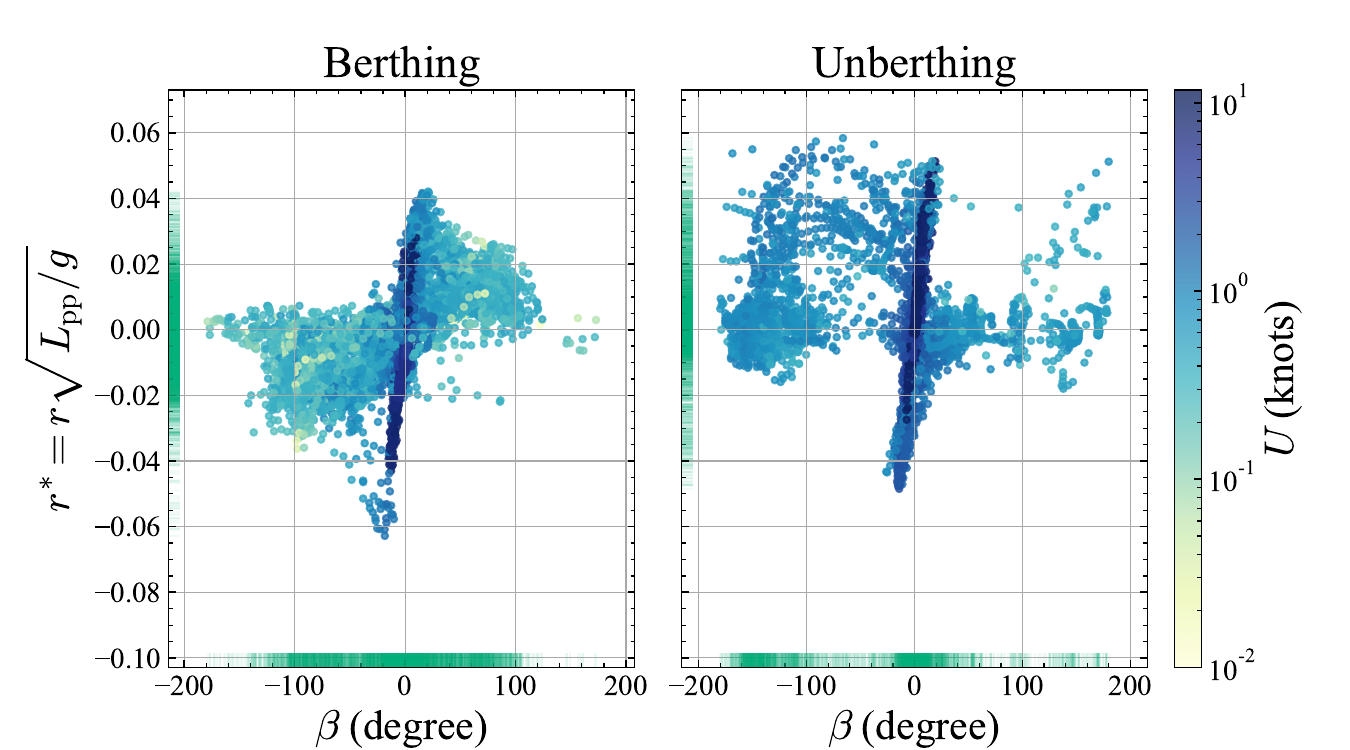}
        \subcaption{All drift angle.}
    \end{minipage} 
    \begin{minipage}[tb]{\linewidth}
        \centering
        \includegraphics[keepaspectratio, width = 0.95\hsize]{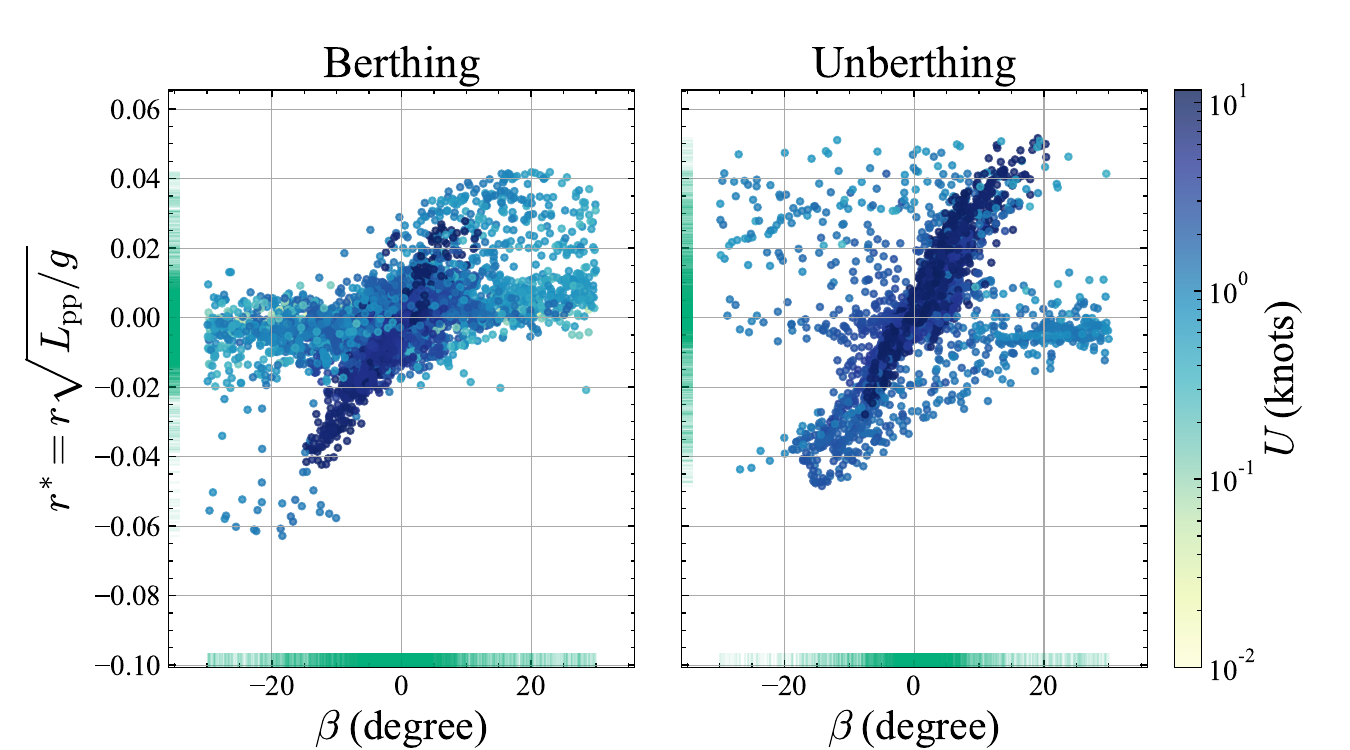}
        \subcaption{$|\beta<30|\degree$ only.}
    \end{minipage} \\

    \caption{Correlation between $\beta$ and  $r^{*}$ for the entire $U$ range.}
    \label{fig:beta_rstar}
\end{figure}

 Additionally, the correlation between $\beta$ and  $r^{*}$  was analyzed using the method proposed by Kose et al. \cite{KOSE1985}, $r^{*} = r\sqrt{\lpp/g}$,  to non-dimensionalize $r$. As illustrated in \cref{fig:beta_rstar}, it is evident that a strong correlation between $\beta$ and $r^{*}$ exists in the high-speed regions. Therefore, it can be confirmed that  $\beta$ and $r$ have a strong positive correlation in the $|\beta|<20\degree$ range, within the speed range close to 10 knots for both berthing and unberthing. 

Moreover, during berthing, a more moderate $\beta - r $ correlation exists for a wide range of $\beta$ below $1$ knot. On the other hand, for unberthing, points where $\beta$ and $r$ have opposite signs are distributed in the second quadrant of the figure. Observation of the unberthing trajectories for each port listed in \cref{app: appendix A} shows that Ports B and D show a right-turning ($r>0$) movement while drifting sternways ($\beta<-\pi$) immediately after the commencement of unberthing.

\begin{figure}[htbp]
    \centering
    \includegraphics[width=0.95\hsize]{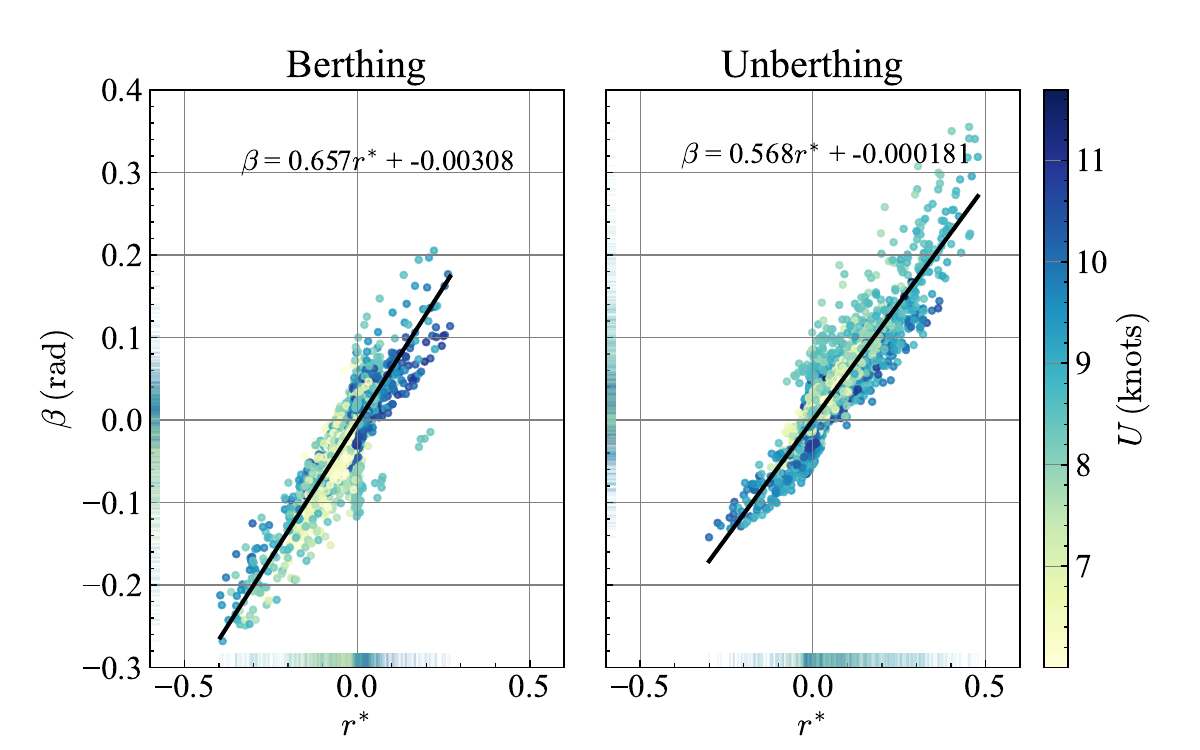}
    \caption{ Correlation between $\beta$ and $r^{*}$ for $U > 6$ knots. }
    \label{fig:r_beta_6ky}
\end{figure}

\subsubsection{Correlation between lateral velocity, $\vm$ and $r$}
Empirical observations indicate that there is a strong $ \vm - r $ correlation for motions at low maneuvering speeds, such as turning and zigzag tests \cite{Abkowitz1980, Sutulo2014a}. It has also been reported that this strong $ \vm - r $ poses significant challenges when estimating coefficients in polynomial-type models such as MMG and Abkowitz models \cite{Hwang1982}. This section aims to ascertain whether the aforementioned strong correlation is also evident in berthing and unberthing maneuvering motions.

\begin{figure}[htbp]
\begin{minipage}[h!]{\linewidth}
    \centering
    \includegraphics[width=0.95\hsize]{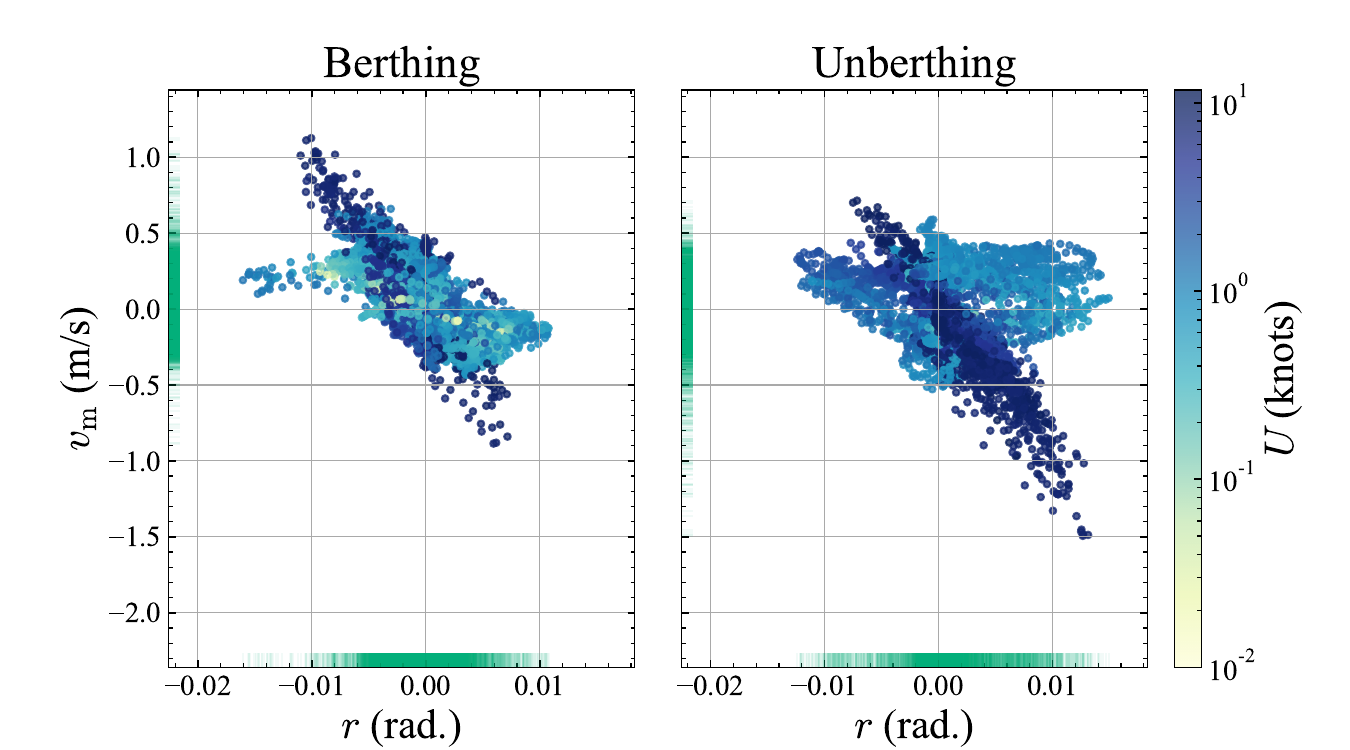}
    \subcaption{All data. The color bar is presented in a log scale.}
    \label{fig:vm_r_all}
\end{minipage}
\begin{minipage}[h!]{\linewidth}
    \centering
    \includegraphics[width=0.95\hsize]{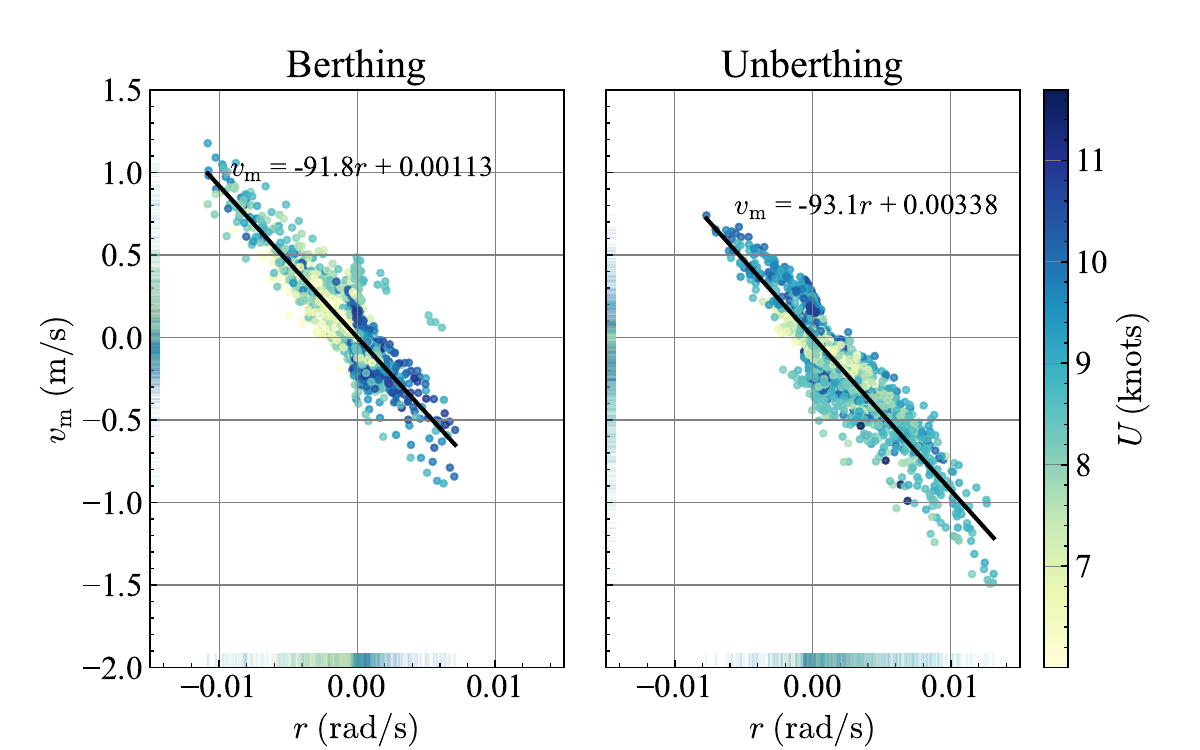}
    \subcaption{$U > 6$ knots only. The color bar is presented on a linear scale.}
    \label{fig:vm_r_6kt}
\end{minipage}
\caption{Correlation between $\vm$ and $r$.}
\label{fig:scatter_vm_r}
\end{figure}

\cref{fig:scatter_vm_r} shows the correlation between $\vm$ and $r$. Similar to the $\beta - r^{*}$  correlation, the $\vm- r$ correlation is strong in the high-speed region and weak in the low-speed region. Consequently, the dataset was divided into high and low-speed categories, using $U = 6$ knots as the threshold. $U=6$ represents approximately half of the highest value of $U$ observed in the dataset. It also corresponds to the speed at which the captain controls the ship to generate a backward force, as described in \cref{sec:rudderangle}. The correlation coefficients of $u,~\vm$, and $r$ are presented in  \cref{tab:corr_uvr}. The correlation coefficients indicate that there is a very strong negative  $\vm  - r$ correlation during both berthing and unberthing at speeds exceeding $6$ knots. Conversely, at speeds below $6$ knots, the correlation is relatively weak during berthing, and nearly absent during unberthing. This trend can be observed in \cref{fig:beta_rstar} and \cref{fig:vm_r_all}. 

Strong $\vm-r$ correlations have been documented at normal speed in both turning and zig-zag tests. Therefore, despite the speed fluctuations that occur during approach and departure maneuvers, when operating at speeds exceeding a certain threshold, such as $U = 6$ knots, the strong $\vm - r $ correlation should be considered in future estimations of the parameters of the berthing and unberthing maneuvering models.

\begin{table}[htbp]
    \caption{Correlation coefficients between $u,~\vm$, and $r$ during berthing and unberthing. }
    \centering
    \begin{minipage}[tb]{\hsize}
    \subcaption{All data.}
        \centering
        \begin{tabular}{cccc|cccc}
        \toprule
            \multicolumn{4}{c}{Berthing}& \multicolumn{4}{c}{Unberthing}\\
            \cmidrule(rl){1-4} \cmidrule(rl){5-8}
            &$u$& $\vm$ & $r$& &$u$& $\vm$ & $r$\\
            $u$ & 1.00 & 0.0662 & -0.108 & $u$ & 1.00 & -0.350 & -0.0218 \\
            $\vm$& & 1.00 & -0.739 & $\vm$& & 1.00 & -0.325 \\
            $r$ & & &   1.00 & $r$ & & &   1.00 \\
            \bottomrule\\\\
        \end{tabular}    
    \end{minipage}\\
    \begin{minipage}[tb]{\hsize}
        \centering
    \subcaption{$U < 6 $ knots.}
        \begin{tabular}{cccc|cccc}
        \toprule
            \multicolumn{4}{c}{Berthing}& \multicolumn{4}{c}{Unberthing}\\
            \cmidrule(rl){1-4} \cmidrule(rl){5-8}
            &$u$& $\vm$ & $r$& &$u$& $\vm$ & $r$\\
            $u$ & 1.00 & 0.0974 & -0.129 & $u$ & 1.00 & -0.285 & -0.186 \\
            $\vm$& & 1.00 & -0.710 & $\vm$& & 1.00 & -0.0107 \\
            $r$ & & &   1.00 & $r$ & & &   1.00 \\
            \bottomrule\\\\
        \end{tabular}  
    \end{minipage}\\
    \begin{minipage}[tb]{\hsize}
        \centering
    \subcaption{$U > 6$ knots.}
        \begin{tabular}{cccc|cccc}
        \toprule
            \multicolumn{4}{c}{Berthing}& \multicolumn{4}{c}{Unberthing}\\
            \cmidrule(rl){1-4} \cmidrule(rl){5-8}
            &$u$& $\vm$ & $r$& &$u$& $\vm$ & $r$\\
            $u$ & 1.00 & -0.271 & 0.299 & $u$ & 1.00 & 0.0188 & 4.59e-4 \\
            $\vm$& & 1.00 & -0.882 & $\vm$& & 1.00 & -0.903 \\
            $r$ & & &   1.00 & $r$ & & &   1.00 \\
            \bottomrule
        \end{tabular}  
    \end{minipage} 
    \label{tab:corr_uvr}
\end{table}
\subsection{Pivot point}\label{sec: pivot}
\begin{figure}[h!]
    \centering
    \includegraphics[width=0.95\hsize]{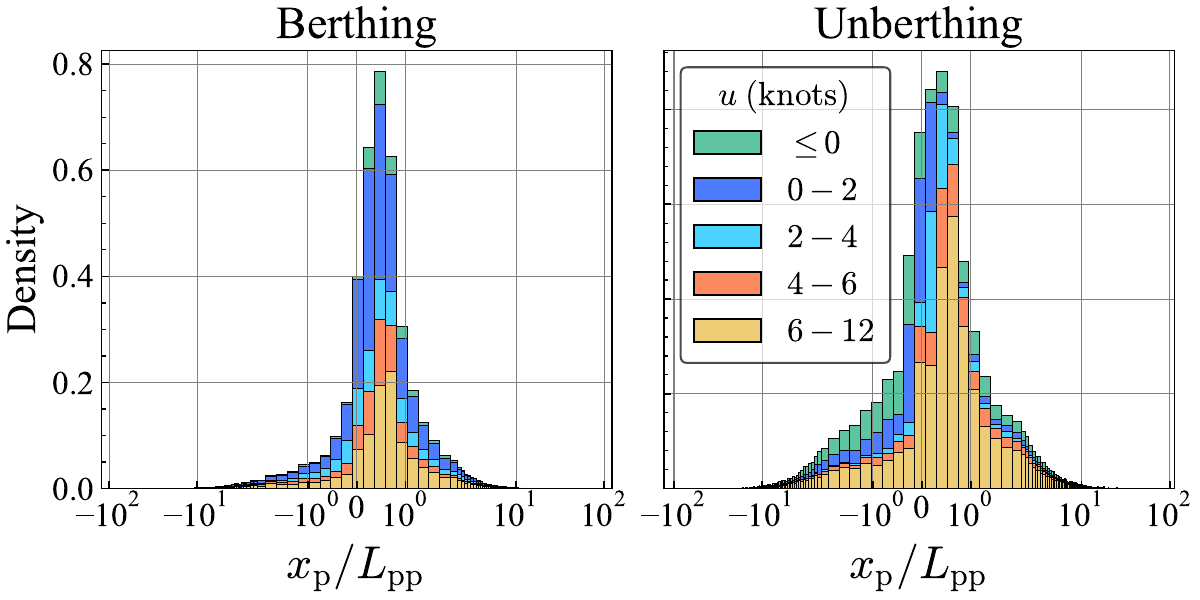}
    \caption{Distribution of $x_{\mathrm{p}}$ location with respect to $u$ during berthing and unberthing.}
    \label{fig:pivot}
\end{figure}

 The pivot point, located along the hull's centerline, is the point about which the ship rotates. This point plays a critical role in determining how the ship responds to rudder and thruster inputs, thereby influencing the ship's turning radius and trajectory. Accurate knowledge of the pivot point's position is essential for precise control of the ship’s movements, which in turn facilitates safer and more efficient berthing and unberthing operations. The position of the pivot point is not fixed; it varies with the ship’s speed, direction, and hull-form\cite{molland2011maritime}. However, it typically lies between one-third and one-sixth of the ship's length, either from the bow if the ship is moving forward or from the aft if the ship is moving astern. At the pivot point, the transverse velocity is zero for $r \neq 0$. 

\begin{equation}\label{eq:pivot1}
    \vm + rx_{\mathrm{p}} = 0
\end{equation}
The position of the pivot point in the $X$-direction, denoted as $x_{\mathrm{p}}$, measured from the midship, is determined by the ratio of  $\vm$ to $r$ as shown in \cref{eq: xp}. \cref{fig:pivot} illustrates the distribution of the pivot point's location during berthing and unberthing.

\begin{equation} \label{eq: xp}
    x_{\mathrm{p}} \equiv -\vm/r
\end{equation}

\subsection{Statistical properties of actuators utilization during berthing and unberthing}
This section presents the statistical analysis of actuators' usage during berthing and unberthing. The subject ship is equipped with a bow thruster and a vectwin rudder system. Accordingly, the analysis focused on the rudder angles and the operational status of the bow thruster as primary parameters of interest. These parameters are integral to understanding the maneuvering dynamics of the ship during berthing and unberthing.
\subsubsection{Vectwin Rudder Angles ($\deltas, \deltap$ )}
\label{sec:rudderangle}

The vectwin rudder system can be steered over a wide range of rudder angles:
\begin{equation}
    \begin{aligned}
        \deltap & = [-105\degree,~35\degree] \\
        \deltas & =[-35\degree, ~105\degree]
    \end{aligned}
\end{equation}

The vectwin rudder system enables the ship to move forward or backward, alter its course, or execute in-situ turns while maintaining a constant propeller speed. These maneuvers can also be enhanced through the utilization of a bow thruster. Additionally, steering can be achieved through the use of a joystick, thereby enabling the operator to execute the aforementioned maneuvers solely by manipulating the joystick \cite{Nabeshima1997_EN}. Moreover, propeller thrust is continuously generated, thereby enabling the generation of steering forces even when the ship is at zero forward speed.

\begin{figure}[htbp]
    \centering
    \includegraphics[width=0.95\hsize]{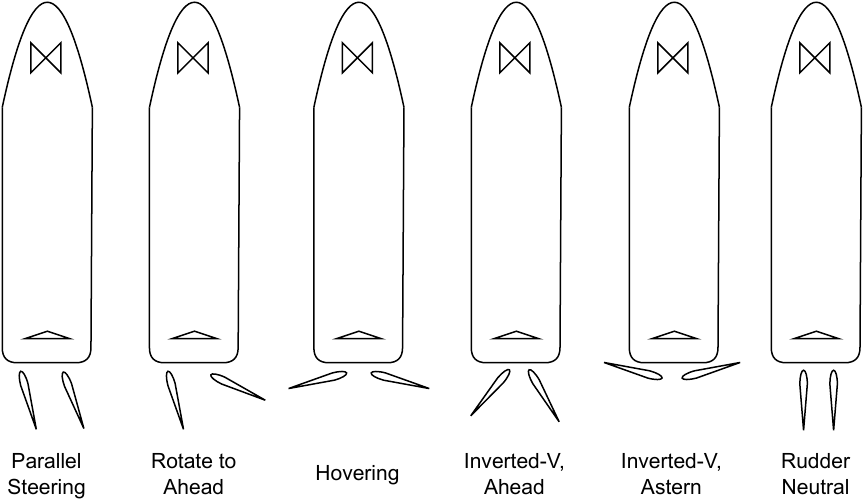}
    \caption{Nominal Vectwin steering mode categories.}
    \label{fig:vectwin_mode}
\end{figure}

In this study, the combinations of rudder angles are categorized into six operational modes, as illustrated in \cref{fig:vectwin_mode}, to represent the operating conditions corresponding to the four propeller quadrants. It is important to note that the rudder angles presented are typical values and may vary based on factors such as propeller rotation speed, load conditions, and other operational parameters. Additionally, fine-tuning the forward/reverse speed can be achieved by slightly tilting the joystick, which is equivalent to adjusting the propeller speed on a conventional single-rudder ship \cite{Nabeshima1997_EN}. A detailed description of each steering mode is provided below.

\begin{description}
    \item[ Mode 1 Parallel steering ] \mbox{}\\
    In this mode, the signs of $\deltas$ and $\deltap$ are identical and the rudder angles are less than $35\degree$, that is, $|\deltasp| \leq 35\degree$. This steering mode is equivalent to that of a conventional single-screw, twin-rudder ship.
    \item[Mode 2 Rotate Ahead] \mbox{}\\
    In this mode, the signs of $\deltas$ and $\deltap$ are identical and one of the rudder angles is greater than $35\degree$. This steering mode is equivalent to the $\lceil$Rotate to port$\rfloor$ or $\lceil$Rotate to stbd$\rfloor$ mode of the vectwin rudder system.
    \item[Mode 3 Hover] \mbox{}\\
    In this mode, the signs of $\deltas$ and $\deltap$ are different and the rudder angles fall within the range of $70\degree \leq |\deltasp| \leq 80\degree $. In this state, the fore and aft forces, lateral forces, and turning moments acting on the hull underwater are smaller, and the ship can be held at a fixed point (hover).
    \item[Mode 4 Inverted-V, Ahead] \mbox{}\\  
    In this mode, the signs of $\deltas$ and $\deltap$ are different and the rudder angles fall within the range of $80\degree$ ($|\deltasp| \leq 80\degree$ ). In this mode, the rudders form a V-shape (inverted "V"), generating a forward force. The magnitude of this forward force can be increased by adjusting the rudder angles closer to zero, thereby moving away from the hover rudder angle.
    \item[Mode 5 Inverted-V, Astern] \mbox{}\\
    In this mode, the signs of $\deltas$ and $\deltap$ are different and the rudder angles are greater than  $90\degree$, that is, ($|\deltasp| \geq 90\degree$ ). In this mode, the rudders form a V-shape (inverted "V"), generating backward force. The magnitude of this backward force can be increased by adjusting the rudder angles further from zero. When combined with a bow thruster, parallel lateral movement is achievable at specific rudder angles.
    \item[Mode 0 Rudder Neutral] \mbox{}\\
    In this mode, both rudders are in a neutral position, with rudder angles maintained below $2\degree$, that is, $|\deltasp| < 2\degree$
     
\end{description}

\begin{figure}[htbp]
    \begin{minipage}[h!]{\linewidth}
        \centering
        \includegraphics[keepaspectratio, width = 0.95\hsize]{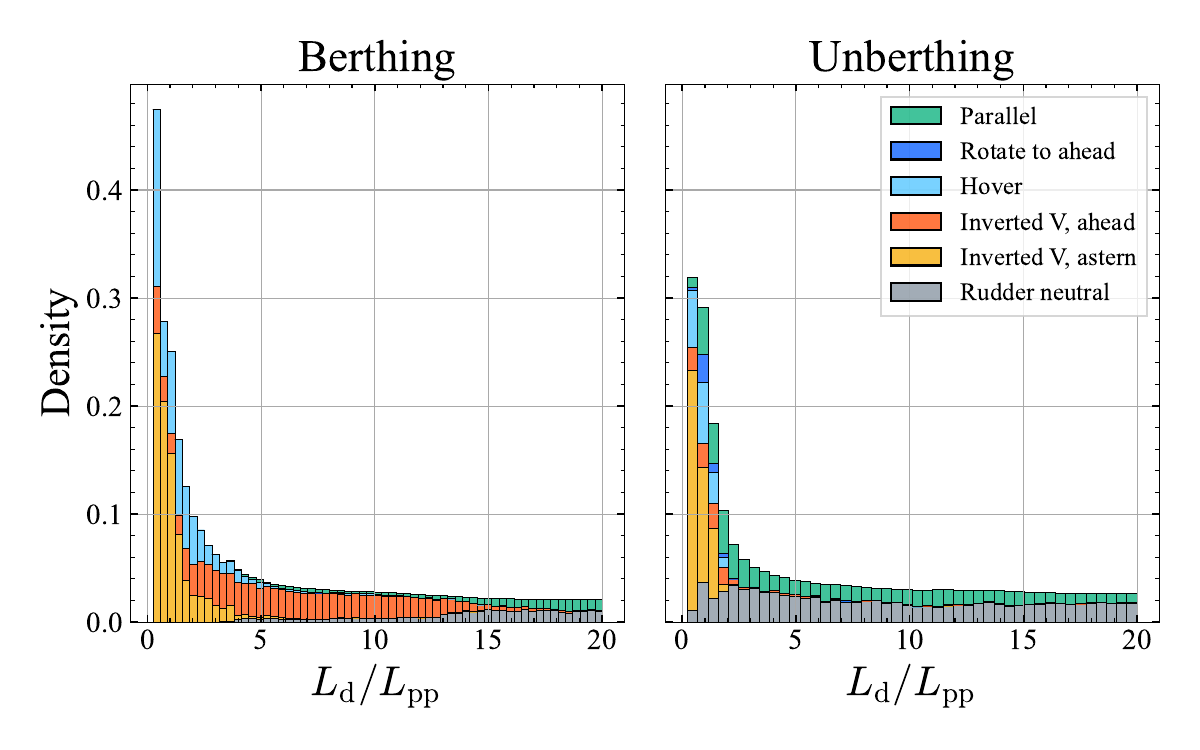}
        \subcaption{All region.}
    \end{minipage} 
    \begin{minipage}[h!]{\linewidth}
        \centering
        \includegraphics[keepaspectratio, width = 0.95\hsize]{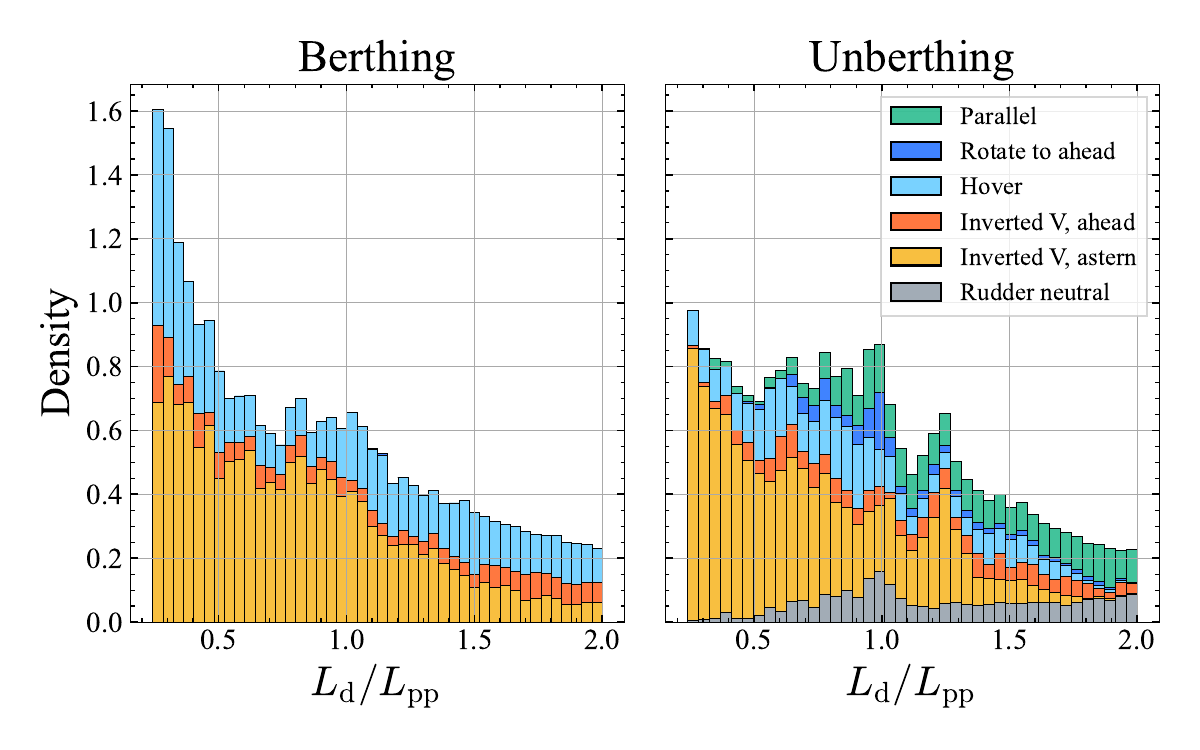}
        \subcaption{Near all region: $\ld \leq 2$.}
    \end{minipage} \\
    \caption{Histogram of steering mode for $\ld$. Histograms are normalized so that the sum of their areas equals 1 in each figures. }
    \label{fig:steering_dist}
\end{figure}

\begin{figure}[htbp]
    \begin{minipage}[h!]{\linewidth}
        \centering
        \includegraphics[keepaspectratio, width = 0.95\hsize]{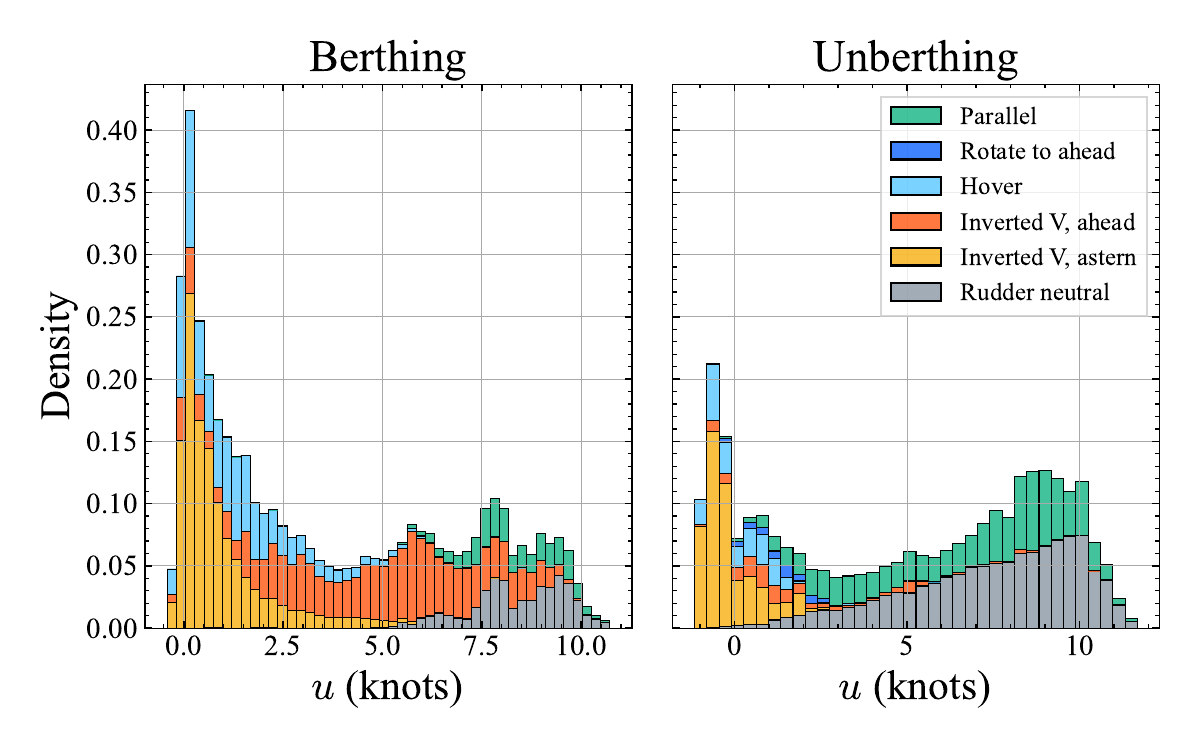}
        \subcaption{All region.}
    \end{minipage} 
    \begin{minipage}[h!]{\linewidth}
        \centering
        \includegraphics[keepaspectratio, width = 0.95\hsize]{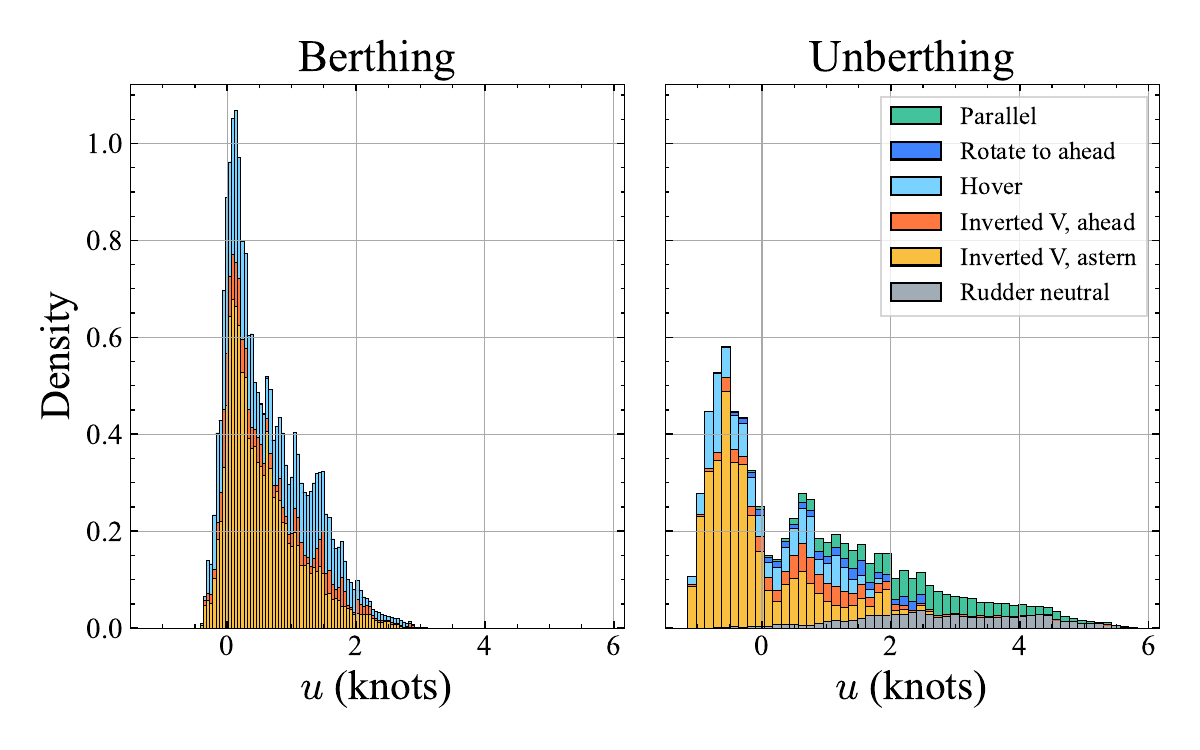}
        \subcaption{Near all region: $\ld \leq 2$.}
    \end{minipage} \\
    \caption{Histogram of steering mode for $u$. Histograms are normalized so that the sum of their areas equals 1 in each figure.}
    \label{fig:steering_u}
\end{figure}

\cref{fig:steering_dist} illustrates the distribution of the steering modes utilization with respect to the ship's distance from the berth, $\ld$. During berthing, Inverted-V Ahead was utilized from a maximum of $\ld/\lpp=20$ to reduce forward force and decelerate the ship. The Inverted-V Astern was utilized from a maximum distance of $\ld/\lpp=5$ to generate a braking force equivalent to propeller reversal. In addition, the Hover and Inverted-V Astern modes were predominantly used during berthing within the $\ld/\lpp <10$ region whereas the Parallel and Rudder-neutral modes were utilized within the $\ld/\lpp > 10 $ region. On the other hand, during unberthing, the Parallel and Rudder-Neutral modes were the most prevalent steering modes, utilized even from distances proximate to the berth.  The Hover and Inverted-V Astern modes were only utilized within the $\ld/\lpp < 3$ region.

Another aspect of interest in maneuvering is the speed at which each steering mode is utilized as illustrated in \cref{fig:steering_u}. It can be observed that, during berthing, the ship started using the Inverted-V Ahead mode from a maximum of $u \approx 10$ knots to decrease the forward force gradually and to generate a braking force from a maximum of $u \approx 6$ knots. On the other hand, during unberthing, the Inverted-V Astern mode was only utilized up to $u\approx3$ knots.

\subsubsection{Bow Thruster}
\begin{figure}[htbp]
        \centering
        \includegraphics[keepaspectratio, width = 0.95\hsize]{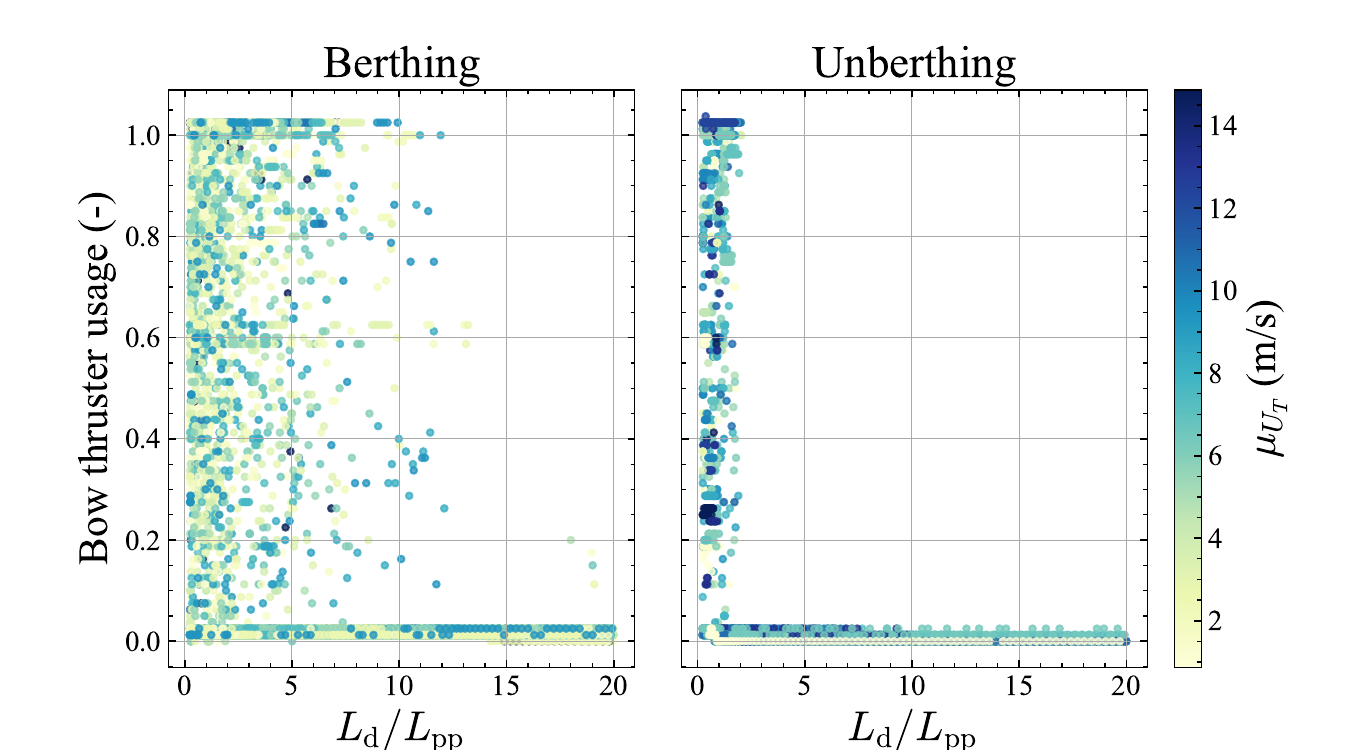}
    \caption{Correlation between the distance of the ship from the berth, $\ld$, and bow thruster usage.}
    \label{fig:dist_bt}
\end{figure}

\begin{figure}[htbp]
        \centering
        \includegraphics[keepaspectratio, width = 0.95\hsize]{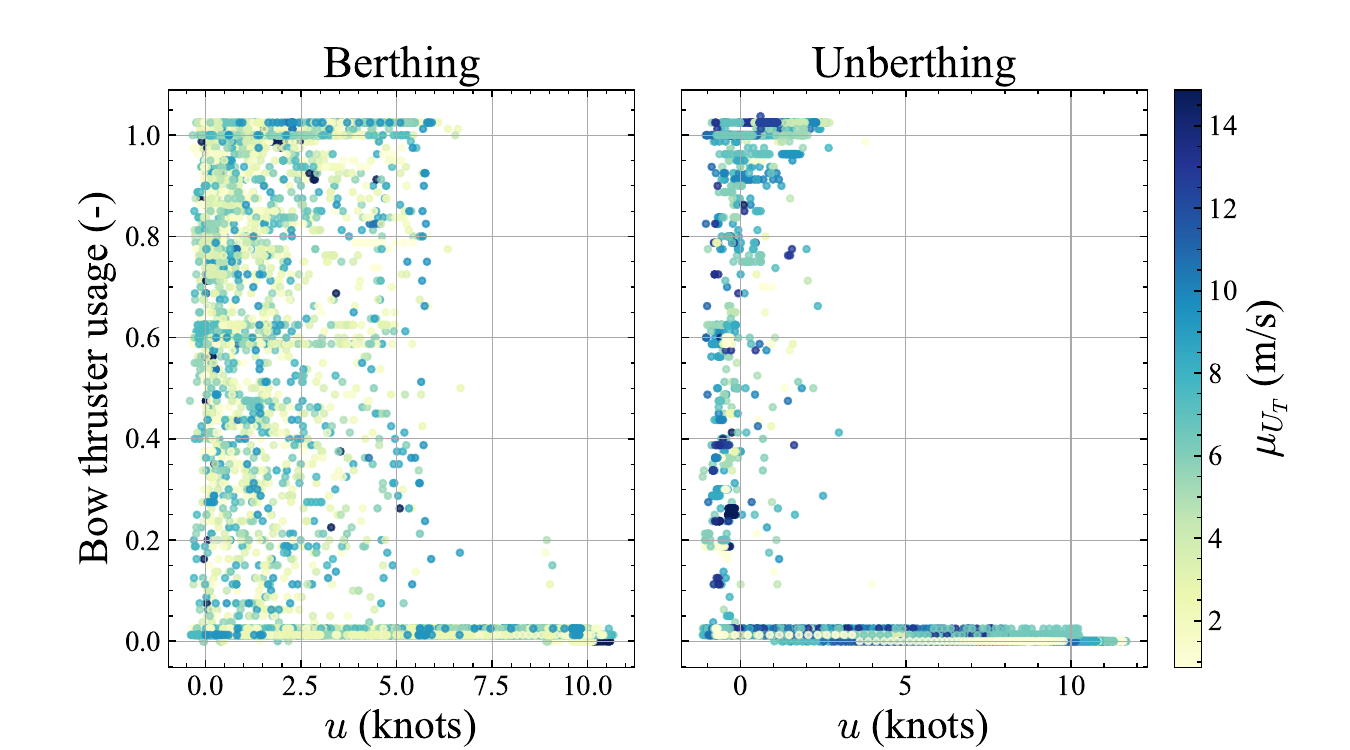}
    \caption{Correlation between the speed of the ship, $u$ and bow thruster usage.}
    \label{fig:u_bt}
\end{figure}

This section presents the characteristics of bow thruster usage. The bow thruster usage $\Tilde{I}\bt$ was normalized using
the commanded current values corresponding to `PORT MAX' and `STBD MAX' operational statuses. Therefore, $\Tilde{I}\bt=0$ is the neutral angle of the bow thruster blade angle and $\Tilde{I}\bt=1$ is the maximum blade angle. \cref{fig:dist_bt} and \cref{fig:u_bt} show the bow thruster usage with respect to distance $\ld$ and forward speed $u$, respectively. Additionally, \cref{fig:dist_bt} and \cref{fig:u_bt} are color-coded according to the average ground wind speed $\Bar{U}_{T}$ for each berthing and unberthing log data.

During unberthing, the bow thruster is utilized predominantly within the region close to the berth, $\ld<2$. On the other hand, during berthing, in the presence of strong winds, $\Bar{U}_{T}>8.0$ m/s, the bow thruster is utilized at distances further from the berth, $\ld>10$. Moreover, bow thruster usage $\Tilde{I}\bt=1$ is observed for  $u \geq 5$ knots in strong wind conditions. This finding is noteworthy, as it is empirically understood that bow thrusters typically do not generate thrust at speeds above $5$ knots. Further investigation is required to elucidate this discrepancy.

\section{Discussion and Limitations}\label{discussion}
In this study, we investigated the maneuvering characteristics of a coastal ship equipped with a vectwin rudder system and a controllable pitch bow thruster. The analysis was conducted using a comprehensive dataset comprising of $153$ operational records collected over approximately one year and encompassing operations in five different ports. The findings of this study, which are summarized below, highlight the key observations and conclusions drawn from the dataset.

\begin{enumerate}[(i)]
    \item Although both berthing and unberthing operations are conducted within confined port environments and share certain similarities, distinct differences exist in their respective maneuvering requirements. As a result, it may be necessary to develop separate considerations for each operation type when designing digital twins, autonomous navigation algorithms, criteria, and guidelines.
    \item The observed trend in speed reduction during the approach maneuvers, particularly during berthing, largely conforms to the speed reduction guidelines proposed by Inoue et al. \cite{Inoue2002_EN}. This statistically and empirically reinforces the applicability of these guidelines in the development of autonomous navigation algorithms and digital twins specific to berthing and unberthing operations. Moreover, the guidelines may be tailored to match the needs of the unberthing maneuvers.
    \item The minimum distance from the midships to the nearest obstacle/wall is approximately equal to the ship length, that is, $d_{\mathrm{ob, min}} \approx \lpp $. This parameter is critical for defining the safe distance necessary for effective collision avoidance in trajectory optimization algorithms and digital twin systems.
    \item Large drift angles, $\beta > 30\degree$ were recorded when the ship was within $\ld / \lpp< 2.0$ range from the berth, with speed generally less than $2$ knots.
    \item The correlation between $\beta - r $ during berthing and unberthing exhibits distinct characteristics compared to the correlation $\beta - r $ during turning and zig-zag tests.
    \item Both berthing and unberthing are associated with large dimensionless yaw rates, $|r^{\prime}|>1$, over the full range of drift angles, with a speed of about $2$ knot or less.
    The dimensionless yaw rate $r^{\prime}_{t}$ observed in this study falls within the range $0 \leq |r^{\prime}_{t}| \leq 1 \times 10 ^ {2}$, which exceeds the range of measurement of typical captive model tests for low-speed maneuvering parameters identification as detailed in \cref{tab: conventional model tests conditions}. This observation highlights the necessity to revise model testing conditions to explore the ship's behavior across a broader range of $r^{\prime}$. Such revisions are expected to enhance low-speed maneuvering models, thereby improving the accuracy of ship dynamics predictions for berthing and unberthing operations.
    
    \begin{table}[htbp]
    \caption{Typical drift angle and yaw rate measurement range in captive model tests for low-speed maneuvering models parameters identification.}
        \centering
        \begin{tabular}{p{1.2cm}p{0.2cm}p{2.0cm}p{2.0cm}}
        \toprule\\
        && Drift angle, $\beta$   & Yaw rate, $r^{\prime}$\\
        \midrule
        Nonaka et al. \cite{nonaka1981experimental} && $0 \sim 180\degree$ & $-0.8 \sim 0.8$ \\
        Kose et al. \cite{kose1984mathematical}&& $0\degree \sim 180\degree$ &$0 \sim 0.15 $\\
        Hirano et al. \cite{hirano1985experimental} & &$-20\degree \sim 20\degree $ & $0.2 \sim 0.8$\\
        Ishibashi et al. \cite{Ishibashi1996}&&$0 \sim 180\degree$ & $0.0 \sim 0.6$\\
        Kobayashi et al. \cite{Kobayashi1998} &&$0 \sim 180\degree$ & $0.0 \sim 0.6$\\
        Yoshimura et al. \cite{Yoshimura2009b}& &$-180\degree \sim 180\degree$& $-2 \sim 2 $ \\
        Yagyu et al. \cite{Yagyu2020} &&$0 \sim 180\degree$& $-2 \sim 2 $\\
        
        \bottomrule
        \end{tabular}
        \label{tab: conventional model tests conditions}
    \end{table}
    \item Through the classification of vectwin rudder operations into six distinct steering modes, the analysis revealed that.:
    \begin{itemize}
        \item During berthing, the Inverted-V steering modes were utilized to gradually reduce forward force from $\ld / \lpp< 20$ and to produce braking force from $\ld/\lpp < 6$
        \item During berthing, the Hover steering mode was predominantly employed within the $\ld/\lpp < 10$ range, whereas parallel and Rudder Neutral steering modes were only utilized in regions away from the berth, $\ld/\lpp > 10 $.
        \item During berthing, the ship was steered such that the forward force was gradually reduced from $u \approx 10$ knots and the braking force was generated from $u \approx 6$ knots.
        \item During unberthing, the Parallel and Rudder-Neutral modes were the most prevalent steering modes, utilized even from distances proximate to the berth.  The Hover and Inverted-V Astern modes were only utilized within the $\ld/\lpp < 3$ region.
        \item During unberthing, the utilization of Inverted-V steering modes can be observed until the ship reaches $u \approx 3$ knots.
        \item  Certain steering modes may be associated with large drift angles. However, an in-depth analysis is required to ascertain this.
    \end{itemize}
    \item The bow thruster was utilized when the ship entered the port under strong wind conditions, particularly at distances $\ld/\lpp \geq 10 $, and at medium speeds of $u \geq 5 $ knots. Further investigation is warranted to determine the effectiveness of the bow thruster under these conditions, as empirical evidence suggests that bow thrusters do not generate significant thrust at speeds exceeding $5$ knots.
\end{enumerate}

Despite the comprehensive analysis presented, certain limitations must be acknowledged. Some of the properties derived may be specific to the subject ship. For instance, the effect of wind on ship motions is highly dependent on the windage area of the ship. Another example is the maneuvering patterns which are dependent on the ship's size and actuators. This specificity may limit the generalizability of the findings across different ship types and configurations. Additionally, although the wind effects were considered, other environmental variables, such as tidal currents and waves, were not fully accounted for. This would specifically limit the information available for the development of digital twins whose accuracy is highly dependent on the accuracy of the empirical data. 

\Add{
}

\section{Conclusion}\label{sec:conclusion}
This study presented a thorough quantitative analysis of the berthing and unberthing maneuvering motions of a full-scale ship, offering a novel empirical validation of previously known qualitative insights.  Furthermore, the findings in this study have significant potential to advance existing maneuvering models, refine captive model test conditions, and contribute to the development and enhancement of digital twins and algorithms for autonomous berthing, unberthing, and in-port maneuvers.  Future research should focus on expanding the dataset to include a broader variety of ships and operational conditions to further validate and enhance the accuracy and robustness of these models and systems.
  
\begin{acknowledgements}
\normalsize{This research was conducted in collaboration with Japan Hamworthy Co. The authors would like to express their deepest gratitude to Japan Hamworthy Co. This work was also supported by a Grant-in-Aid for Scientific Research from the Japan Society for Promotion of Science (JSPS KAKENHI Grant Number 22H01701).}
\end{acknowledgements}

%
\section*{Conflict of interest}
The authors declare that they have no conflict of interest.

\appendix 
\def\thesection{ \Alph{section}\;}
\section{Example of berthing and unberthing at each port.} 
\label{app: appendix A}
\begin{figure*}[htbp]
        \begin{minipage}[h!]{\linewidth}
            \centering
            \includegraphics[width=0.70\linewidth]{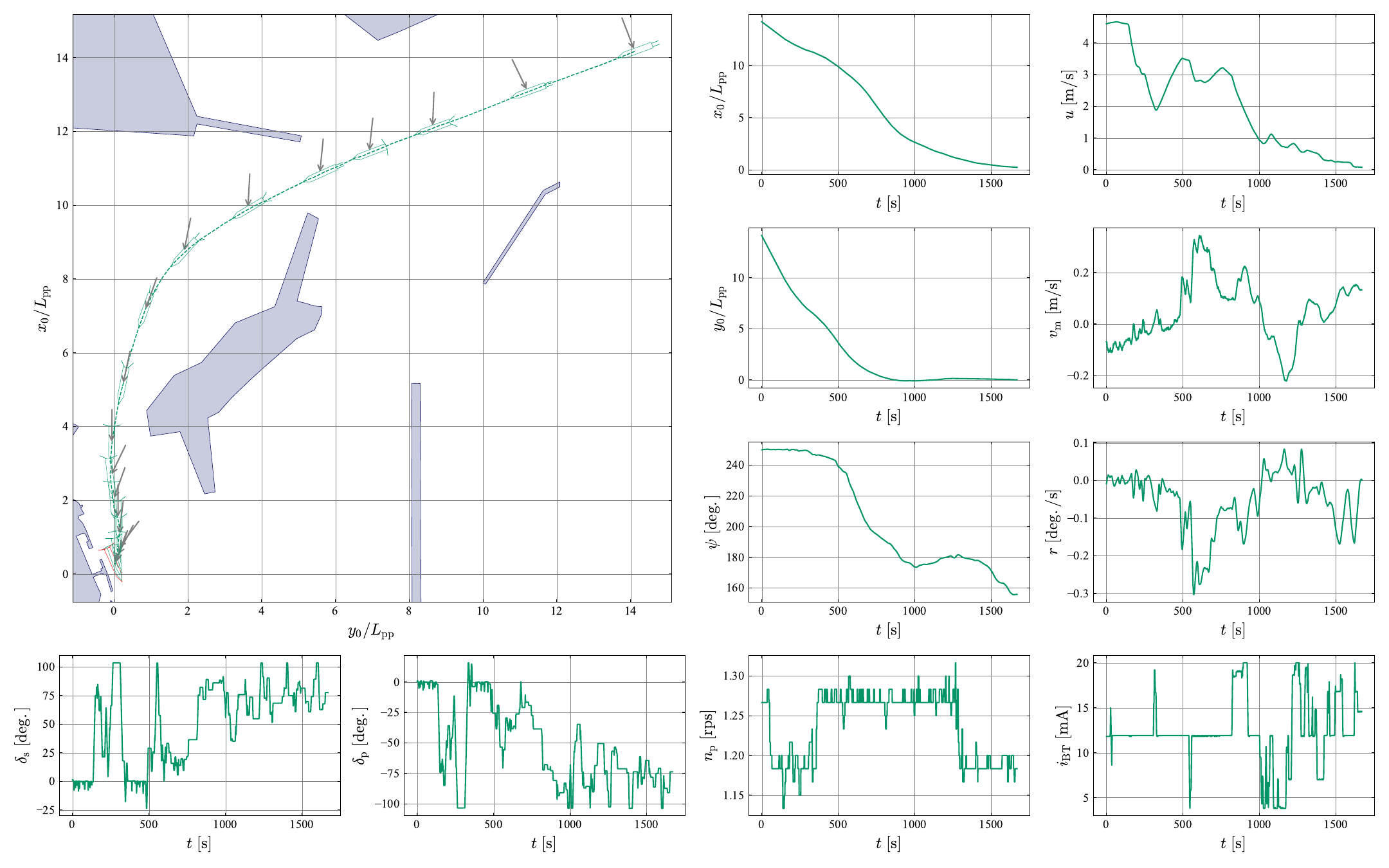}
            \subcaption{Berthing at port A.}
            \label{fig:in_portA}
        \end{minipage}\\
        \begin{minipage}[h!]{\linewidth}
            \centering
            \includegraphics[width=0.70\linewidth]{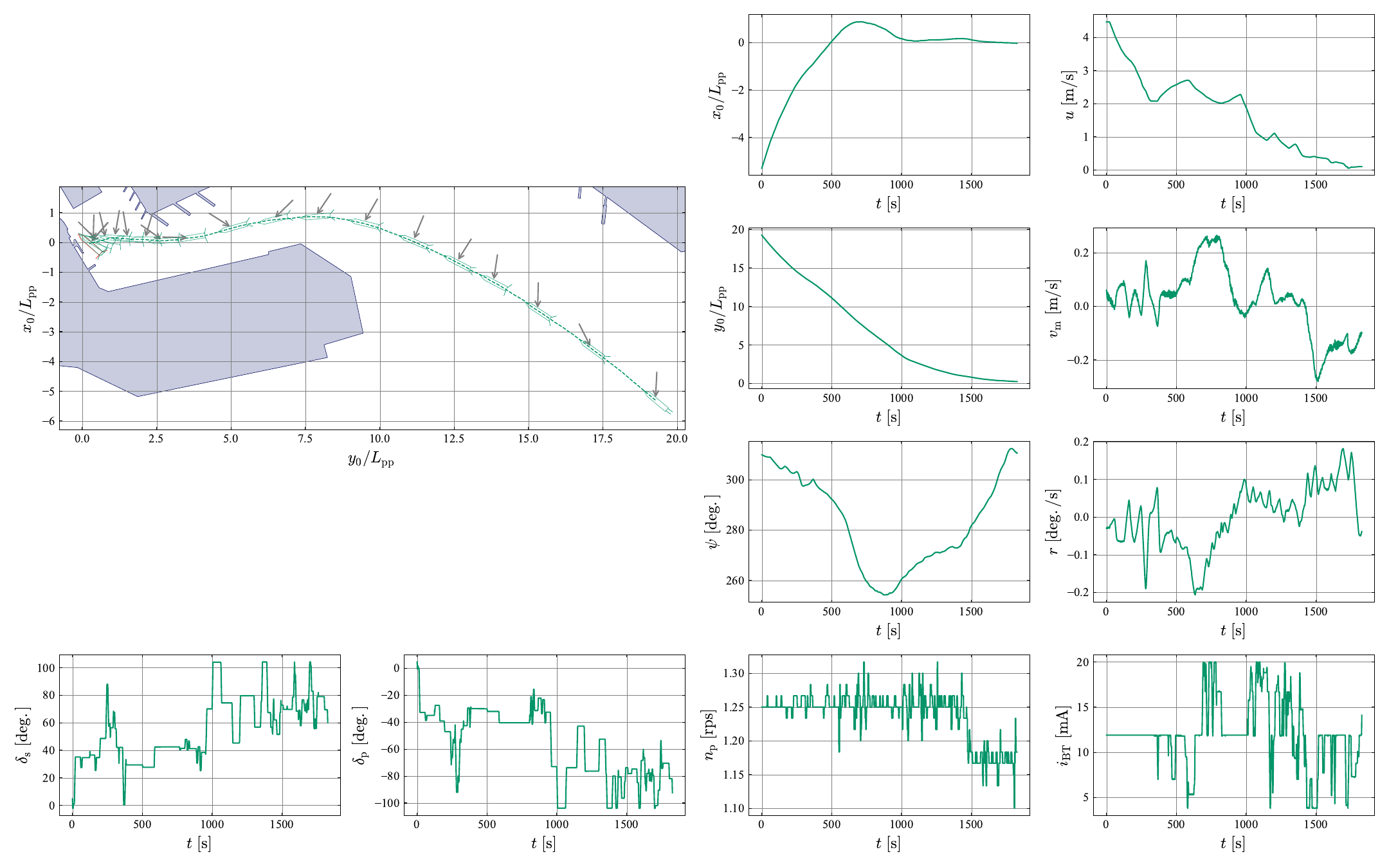}
            \subcaption{Berthing at port B.}
        \label{fig:in_portB}
        \end{minipage} \\
        \begin{minipage}[h!]{\linewidth}
            \centering
            \includegraphics[width=0.70\linewidth]{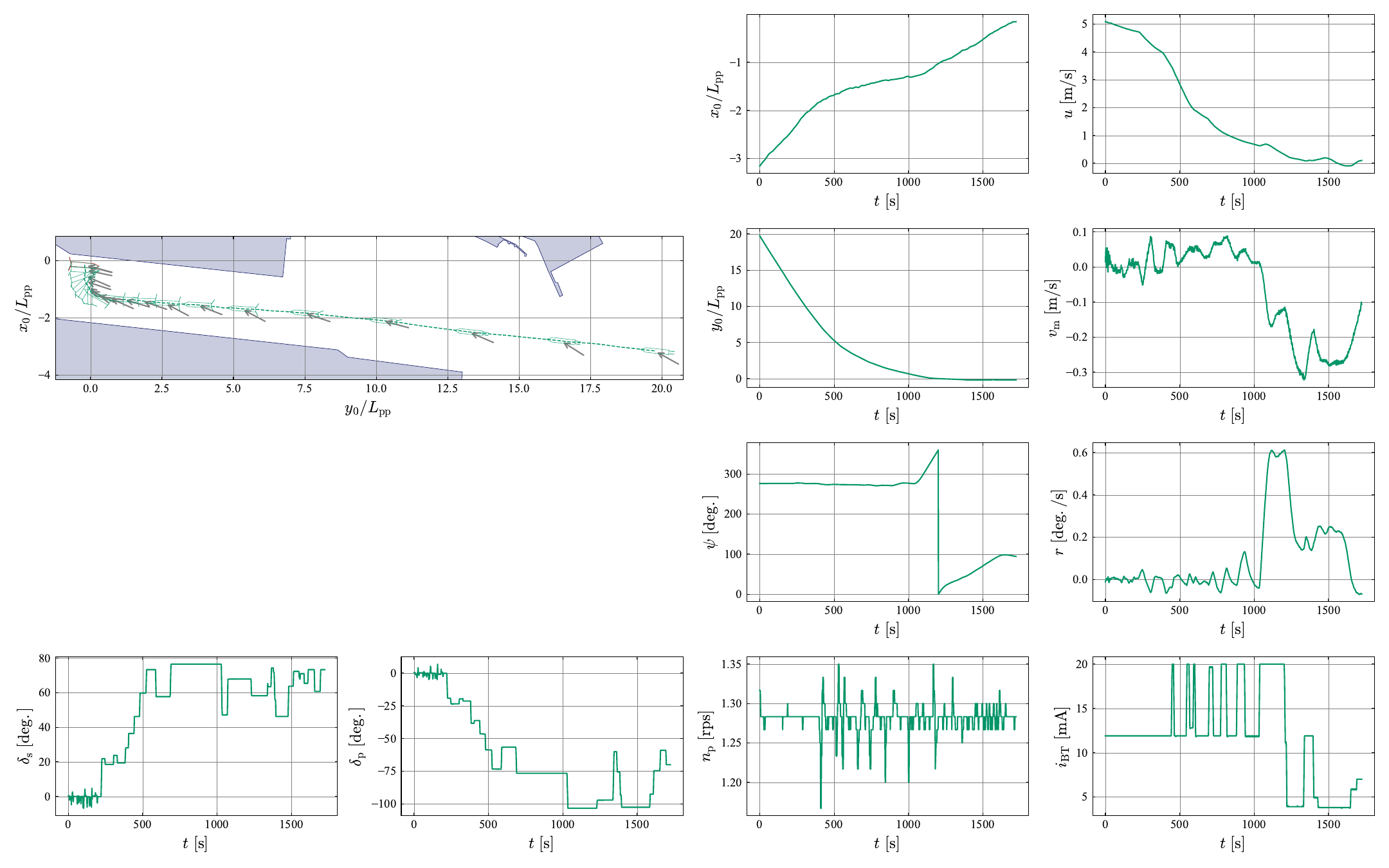}
            \subcaption{Berthing at port C.}
        \label{fig:in_portC}
        \end{minipage} 
    \caption{Example of berthing log data at port A, B, and C.}
    \label{fig:exampleABCin}
\end{figure*}

\begin{figure*}[htbp]
        \begin{minipage}[h!]{\linewidth}
            \centering
            \includegraphics[width=0.70\linewidth]{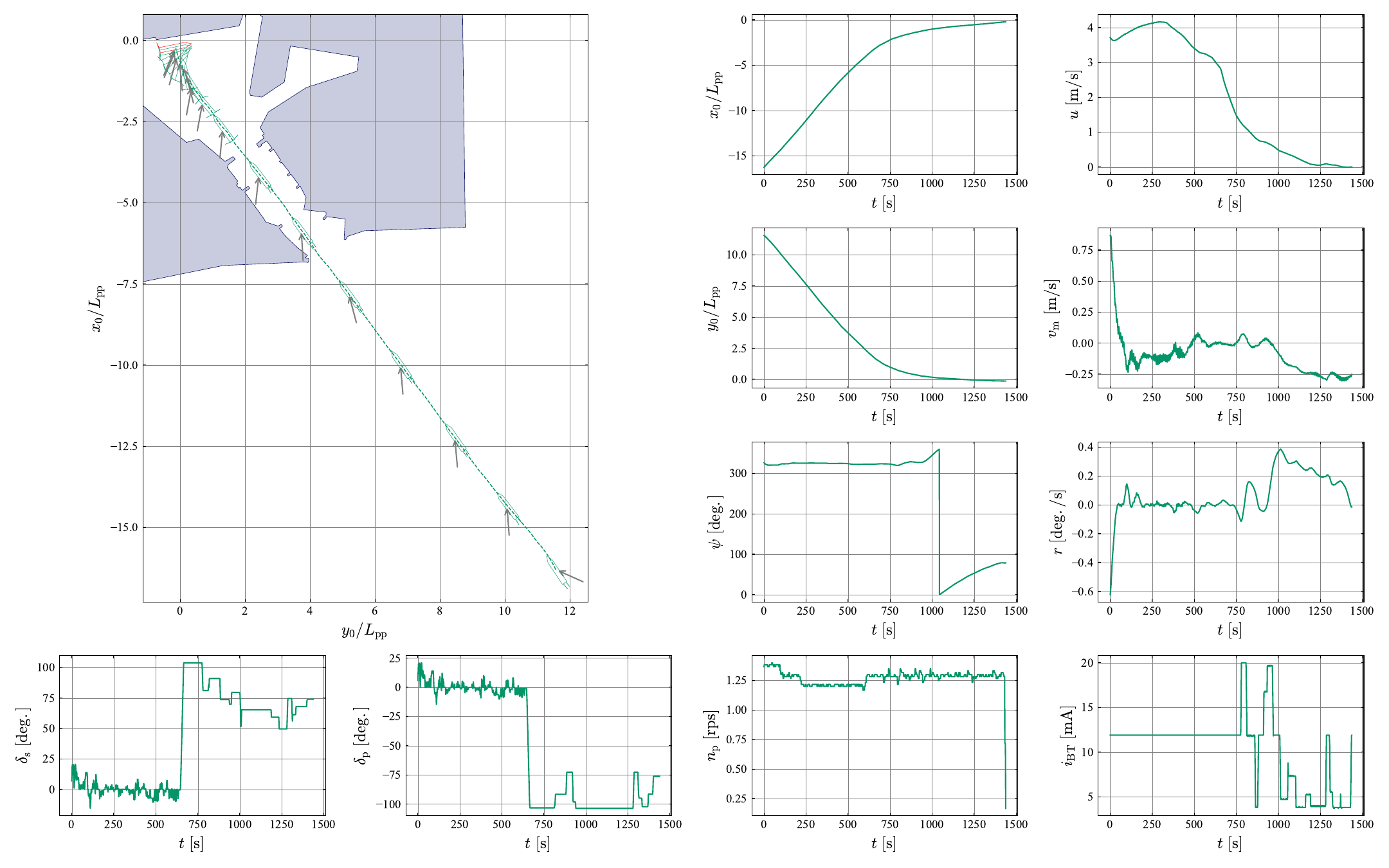}
            \subcaption{Berthing at port D.}
        \label{fig:in_portD}
        \end{minipage}  \\
    \begin{minipage}[h!]{\linewidth}
            \centering
            \includegraphics[width=0.70\linewidth]{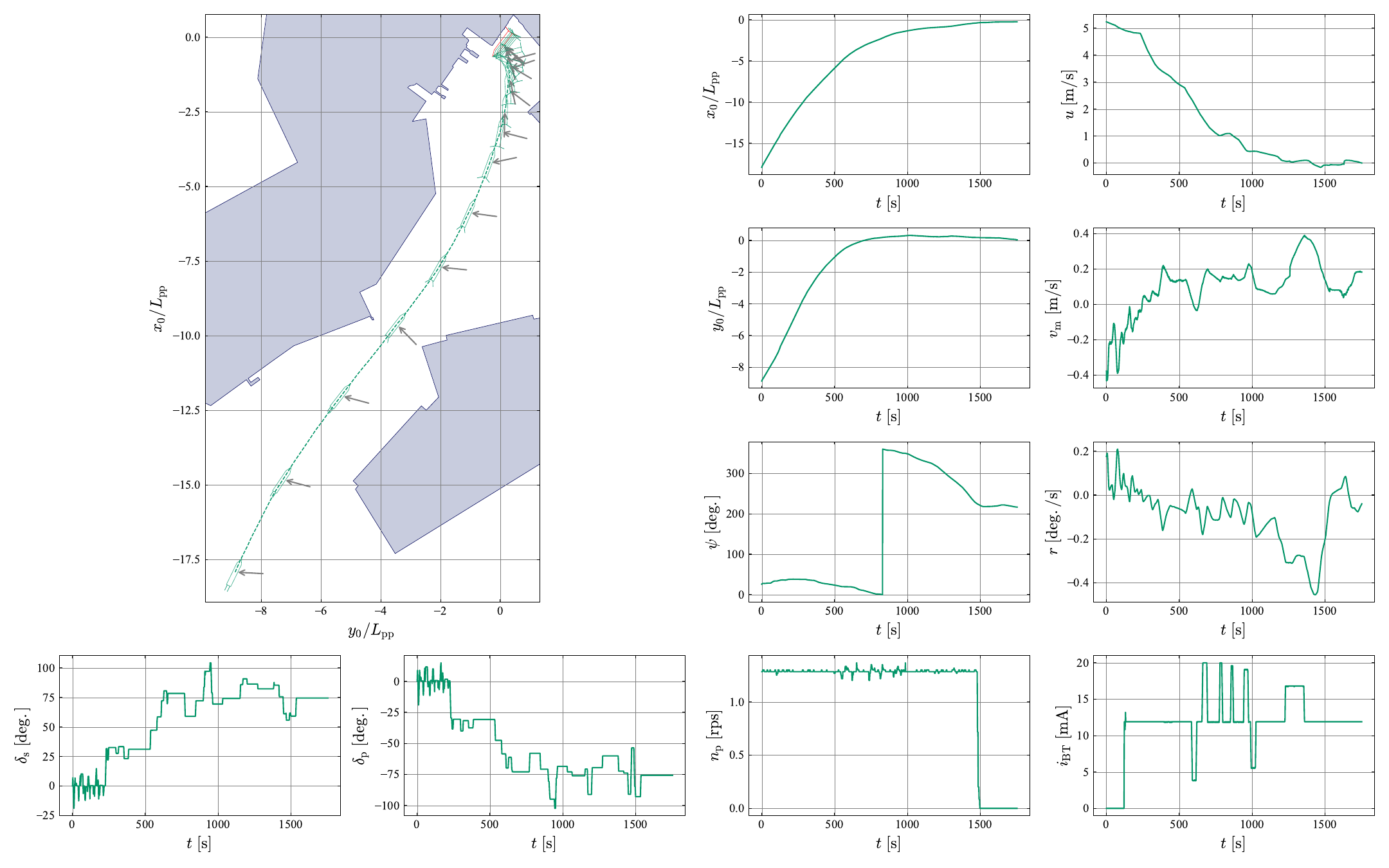}
            \subcaption{Berthing at port E.}
        \label{fig:in_portE}
    \end{minipage} \\
    \caption{Example of berthing log data at port D and E-1.}
    \label{fig:exampleDEin}
\end{figure*}

\begin{figure*}[htbp]
        \begin{minipage}[h!]{\linewidth}
            \centering
            \includegraphics[width=0.70\linewidth]{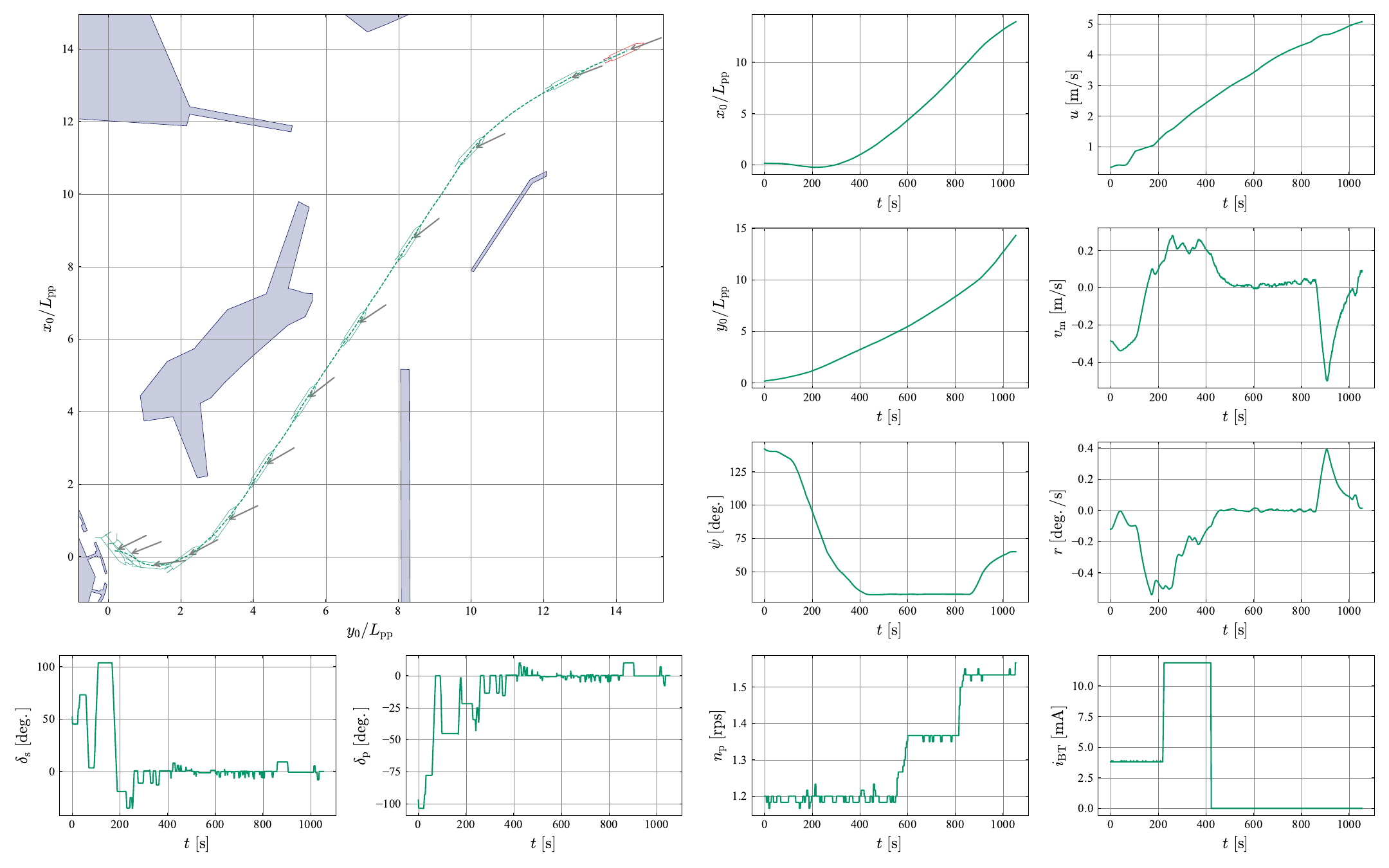}
            \subcaption{Unerthing at port A.}
            \label{fig:おout_portA}
        \end{minipage}\\
                \begin{minipage}[b]{\linewidth}
            \centering
            \includegraphics[width=0.70\linewidth]{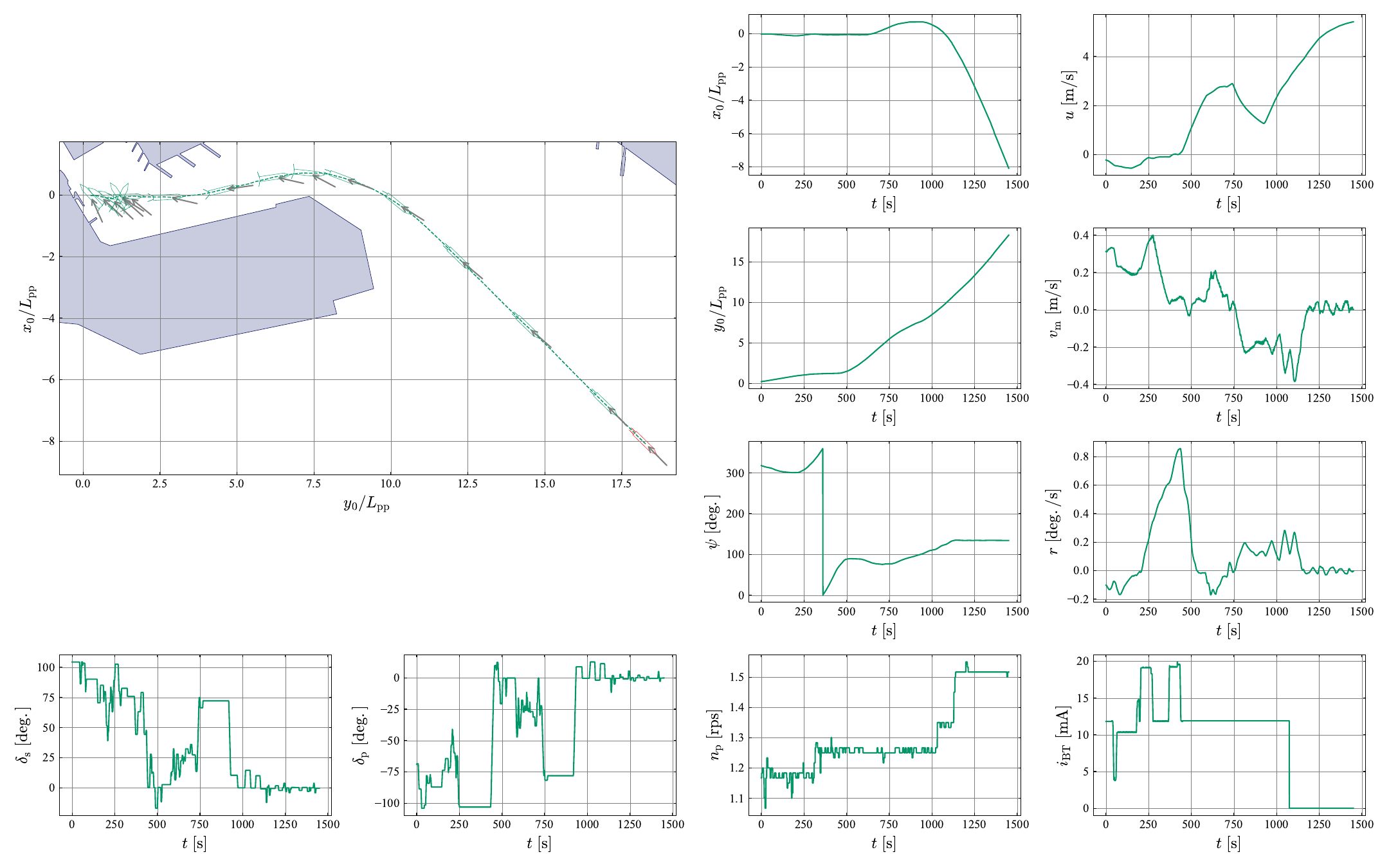}
            \subcaption{Unerthing at port B.}
        \label{fig:out_portB}
        \end{minipage} \\
        \begin{minipage}[h!]{\linewidth}
            \centering
            \includegraphics[width=0.70\linewidth]{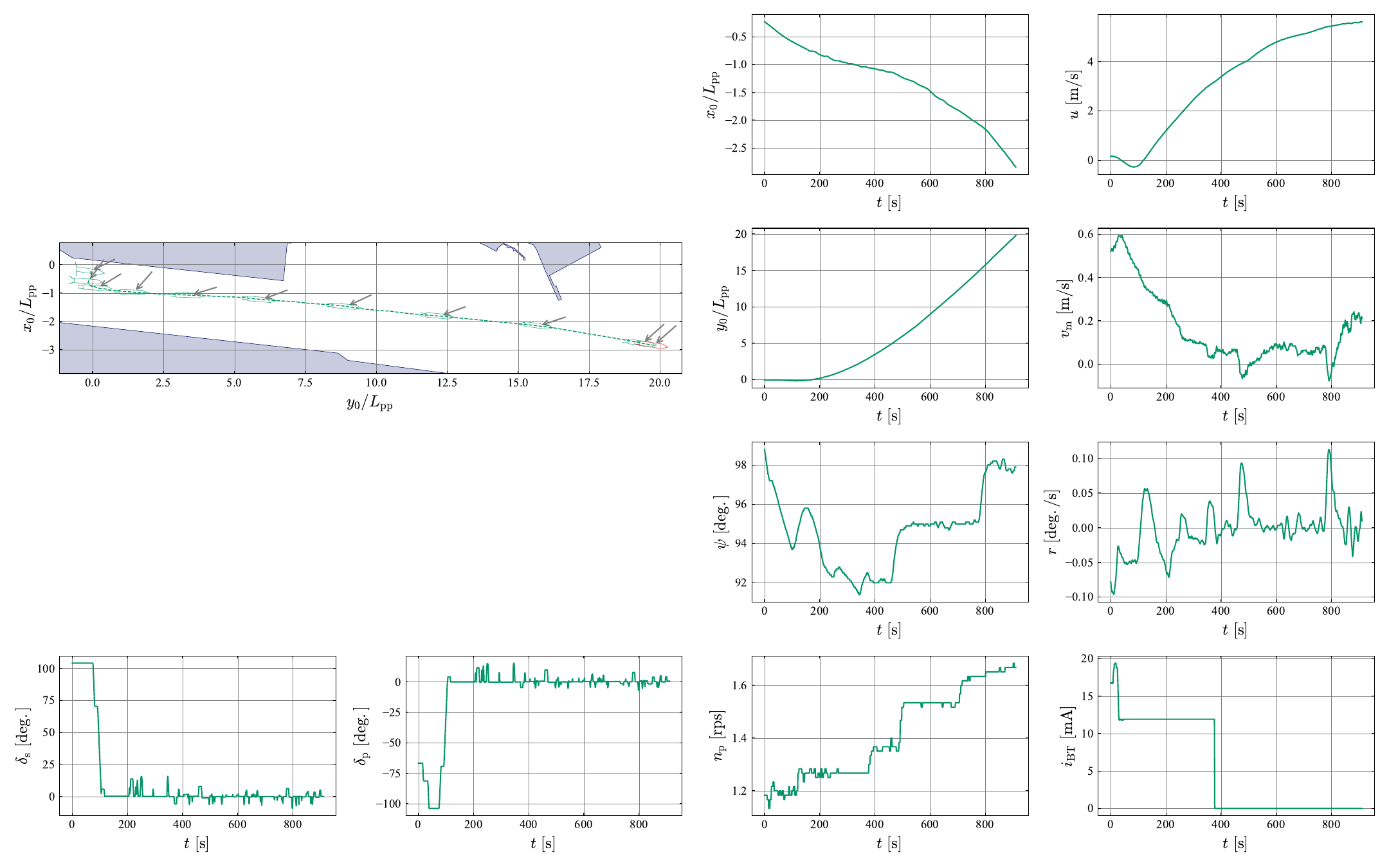}
            \subcaption{Unberthing at port C.}
        \label{fig:out_portC}
        \end{minipage} \\
    \caption{Example of unberthing log data at port A, B, and C.}
    \label{fig:exampleABCout}
\end{figure*}

\begin{figure*}[htbp]
    \begin{minipage}[h!]{\linewidth}
            \centering
            \includegraphics[width=0.70\linewidth]{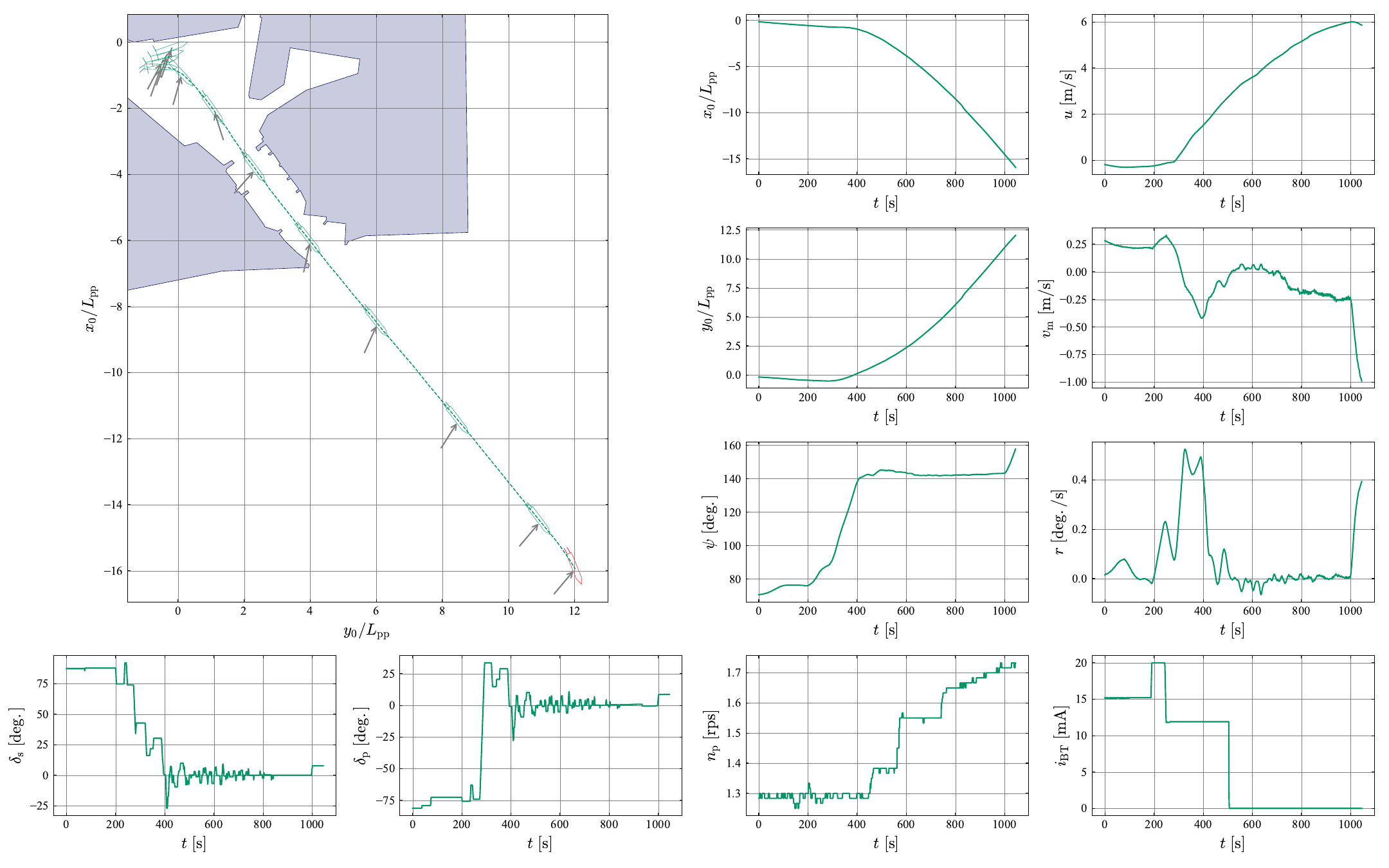}
            \subcaption{Unberthing at port D.}
        \label{fig:out_portD}
    \end{minipage} \\
    \begin{minipage}[h!]{\linewidth}
            \centering
            \includegraphics[width=0.70\linewidth]{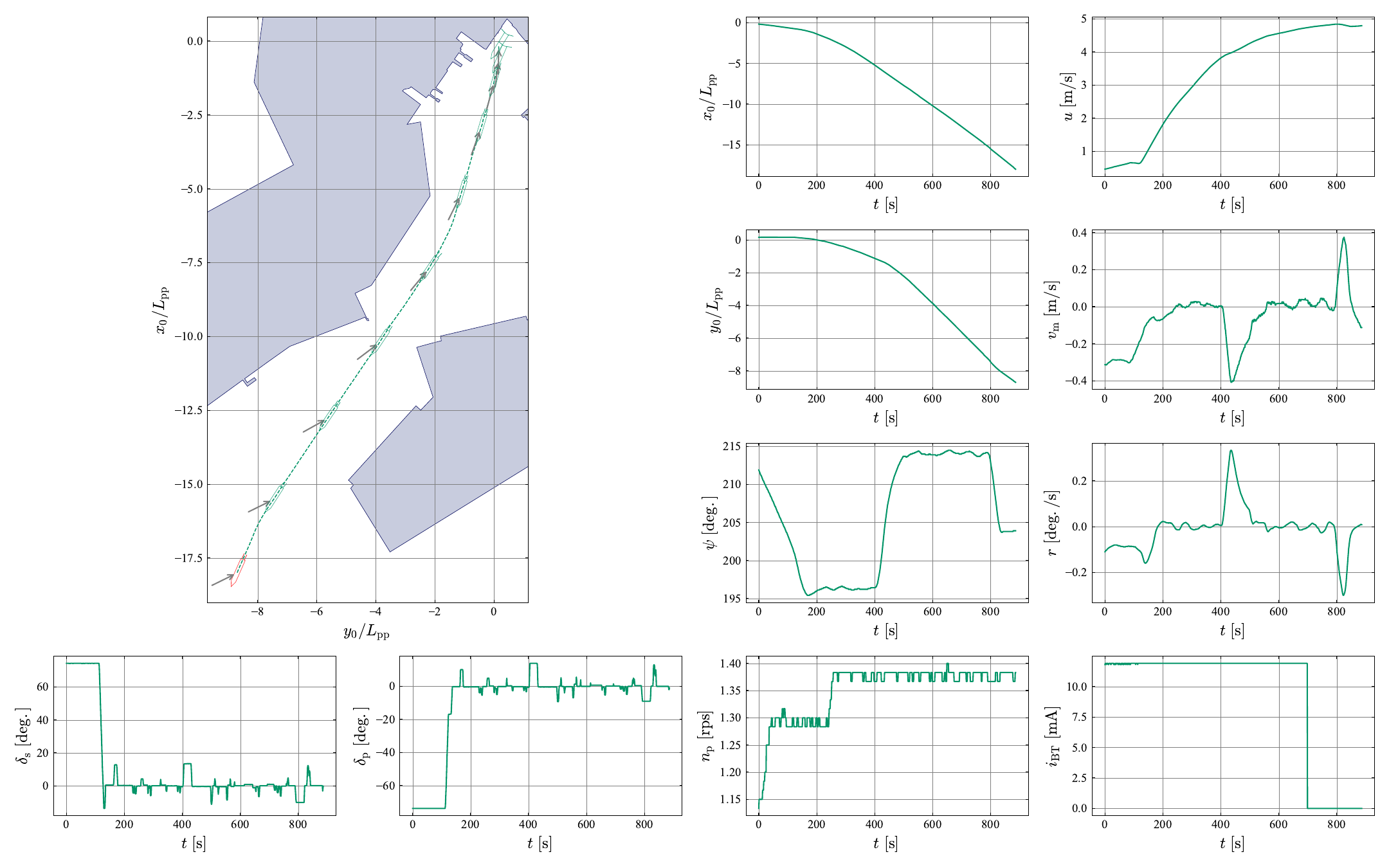}
            \subcaption{Unberthing at port E-1.}
        \label{fig:out_portE}
    \end{minipage} 
    \caption{Example of unberthing time histories of port D and E.}
    \label{fig:exampleDEout}
\end{figure*}

\cref{fig:exampleABCin} and \cref{fig:exampleDEin} show examples of berthing at ports A - E. Similarly, \cref{fig:exampleABCout} and \cref{fig:exampleDEout} show examples of unberthing at ports A - E.

\bibliographystyle{spphys}       
\bibliography{main.bib}   
\end{document}